\documentclass[manuscript]{acmart}

\AtBeginDocument{%
  \providecommand\BibTeX{{%
    \normalfont B\kern-0.5em{\scshape i\kern-0.25em b}\kern-0.8em\TeX}}}

\copyrightyear{2022} 
\acmYear{2022} 
\setcopyright{rightsretained} 
\acmConference[EAAMO '22]{Equity and Access in Algorithms, Mechanisms, and Optimization}{October 6--9, 2022}{Arlington, VA, USA}
\acmBooktitle{Equity and Access in Algorithms, Mechanisms, and Optimization (EAAMO '22), October 6--9, 2022, Arlington, VA, USA}
\acmDOI{10.1145/3551624.3555300}
\acmISBN{978-1-4503-9477-2/22/10}

\usepackage[ruled,vlined]{algorithm2e}

\usepackage{amsmath, amsthm,amsfonts, amssymb,epstopdf,color,soul}

\usepackage{hyperref}
\usepackage{geometry}%
\usepackage{graphicx}
\usepackage{subcaption}
\usepackage{bm}
\usepackage[shortlabels]{enumitem}
\usepackage[titletoc,page]{appendix}
\usepackage{float}
\usepackage{natbib}
\usepackage{bbold}
\DeclareMathAlphabet{\mathbb}{U}{bbold}{m}{n}
\usepackage{newtxmath}
\usepackage{soul}

\usepackage[ruled,vlined]{algorithm2e}

\usepackage{amsmath, amsthm,amsfonts, amssymb,epstopdf,color,soul}

\usepackage{hyperref}
\usepackage{graphicx}
\usepackage{subcaption}
\usepackage{bm}
\usepackage[shortlabels]{enumitem}
\usepackage[titletoc,page]{appendix}
\usepackage{float}
\usepackage{newtxmath}
\usepackage{soul}
\usepackage{tabularx}

\begin{document}

\title{Mathematically Quantifying Non-responsiveness of the \\2021 Georgia Congressional Districting Plan}

\author{Zhanzhan Zhao}
\authornotemark[2]
\email{zhanzhanzhao@gatech.edu}
\affiliation{
  \institution{Georgia Institute of Technology}
   \country{USA}
}

\author{Cyrus Hettle}
\authornote{Second first author.}
\email{chettle@gatech.edu}
\affiliation{
  \institution{Georgia Institute of Technology}
  \department{School of Mathematics}
     \streetaddress{North Avenue}
  \city{Atlanta}
  \state{Georgia}
  \country{USA}
  \postcode{30332}
}

\author{Swati Gupta}
\email{swatig@gatech.edu}
\affiliation{
  \institution{Georgia Institute of Technology}
   \country{USA}}

\author{Jonathan C. Mattingly}
\email{jonm@math.duke.edu}
\affiliation{%
  \institution{Duke University}
   \country{USA}}

\author{Dana Randall}
\email{randall@cc.gatech.edu}
\affiliation{%
  \institution{Georgia Institute of Technology}
   \country{USA}}

\author{Gregory J. Herschlag}
\authornote{Corresponding authors. Emails: zhanzhanzhao@gatech.edu, gjh@math.duke.edu. \\ This work was supported by NSF Award CCF-2106687, CCF-1733812, CMMI-2112533 and  DMS-1613337.}
\email{gjh@math.duke.edu}
\affiliation{%
  \institution{Duke University}
   \country{USA}}

\renewcommand{\shortauthors}{Zhao, et al.}

\begin{abstract} To audit political district maps for partisan gerrymandering, one may determine a baseline for the expected distribution of partisan outcomes by sampling an ensemble of maps.  
One approach to sampling is to use redistricting policy as a guide to precisely codify preferences between maps.  Such preferences give rise to a probability distribution on the space of redistricting plans, and Metropolis-Hastings methods allow one to sample ensembles of maps from the specified distribution.
Although these approaches have nice theoretical properties and have successfully detected gerrymandering in legal settings, sampling from commonly-used policy-driven distributions is often computationally difficult. As of yet, there is no algorithm that can be used off-the-shelf for checking maps under generic redistricting criteria. In this work, we mitigate the computational challenges in a Metropolized-sampling technique through a parallel tempering method combined with ReCom\cite{deford2019recombination} and, for the first time, validate that such techniques are effective on these problems at the scale of statewide precinct graphs for more policy informed measures. We develop these improvements through the first case study of district plans in Georgia. Our analysis projects that any election in Georgia will reliably elect 9 Republicans and 5 Democrats under the enacted plan. This result is largely fixed even as public opinion shifts toward either party and the partisan outcome of the enacted plan does not respond to the will of the people. Only 0.12\% of the $\sim$160K plans in our ensemble were similarly non-responsive.
\end{abstract}



\maketitle

\section{Introduction}
Gerrymandering is the process of manipulating political districts either to amplify the influence of a political group or suppress the representation of various demographic groups. Over recent redistricting cycles, mathematicians, political scientists, and lawyers have begun to develop effective methodologies to understand and uncover the intent and effects of gerrymandered districts \cite{mattingly2014redistricting, fifield2015, Chikina_Frieze_Pegden_2017, deford2019redistricting, Herschlag20, becker2021computational}. The basic idea behind these methods is to compare a redistricting plan to a large collection of neutrally drawn alternative plans that comply with preferences motivated by legal and policy considerations. In this work, our main goal is to analyze gerrymandering in the 2021 Georgia congressional districts.  We also endeavour to refine and codify the intellectual framework one ideally would use in such analyses and analyze the plans with the Georgia congressional redistricting criteria.

A key approach in detecting gerrymandering that has been upheld in various state courts \cite{2018league,2019harper,2019lewis,2022harper} is as follows: one uses non-partisan criteria reflected by a plan of interest, $M$, to draw a collection of plans, referred to as an {\it ensemble}, from a specified distribution, and then checks whether $M$ is an outlier with respect to partisan properties, such as election outcomes. Creating a collection of plans reflecting the non-partisan properties of $M$ is necessary to be able to directly compare the ensemble to the enacted plan; for example, if the ensemble has districts that are significantly less compact than the enacted plan, one could not assess if differences in partisan behavior came from explicit manipulation or were necessitated by the increased compactness.  Therefore, one may wish to ``tune" (i.e., control) the average compactness\footnote{In the current literature, there is an ongoing debate over which measure of compactness is appropriate in redistricting; the debate includes using spanning trees, ``cut" edges \cite{deford2019recombination}, ML/human learning approaches \cite{kaufman2021measure}, and more traditional measures such as Reock and Polsby-Popper. Despite known issues with the traditional measures \cite{barnes2021gerrymandering}, they are still most often used in the practice of redistricting. This is partially because nearly all measures have known issues. In our case study, we will use the Polsby-Popper measure of compactness.} of the districts in the ensemble of plans to align with the compactness of~$M$. 

In generating ensembles, Metropolized-sampling  approaches transparently allow one to codify preferences between plans in a policy-driven framework, providing a flexible and explicit distribution on the space of plans. However, sampling from many commonly-used policy-driven distributions is computationally difficult and, as of yet, there is no algorithm that can be used off-the-shelf for plans in generic political environments. This is due to barriers in some sampling procedures that can cause extremely slow convergence rates, as well as using proposal chains or generative methods that are close to singular with respect to the desired measure.

Although currently there is no federal prohibition on partisan gerrymandering, the state court cases mentioned above have overturned enacted plans based on these methods. Furthermore, these methods provide a robust and solid footing in how to think about and understand extreme partisanship in redistricting which can serve to inform public debate and policy surrounding redistricting practices.

The key contribution of our work is to mitigate the challenges with sampling from policy-driven distributions via a parallel tempering scheme; parallel tempering is a standard tool to bridge target measures with measures that are easier to sample from, however it does not always work in practice.  It has been used previously in redistricting on diffusive boundary methods \cite{fifield2015,Fifield_Higgins_Imai_Tarr_2020}, however these methods have not been demonstrated to scale to the size of statewide graphs. Here, we employ parallel tempering with recombination; to our knowledge, this is the first time these two methods have been combined and we demonstrate their efficacy on statewide graphs for particular measures. We implement our algorithm via a case study of detection of gerrymandering in the 2021 enacted congressional redistricting plan in Georgia. We summarize our main findings below:

\begin{enumerate}
    
    \item {\bf Accelerated Sampling:} In sampling redistricting plans on Georgia, we tune a distribution to sample from an ensemble of plans that have a comparable level of compactness as the 2021 enacted plan. 
    We bridge algorithmically accessible measures to policy-informed measures via parallel tempering using Metropolized ReCom chains. Although theoretically sampling from this distribution is feasible, in practice, using single chains we find that these show no signs of mixing after days of run time. Using the resulting ensemble, we compare the partisan properties of the ensemble with those of the 2021 enacted plan as explained next.
	\item {\bf Non-responsiveness:} We find that the 2021 plan will likely be highly {\it non-responsive} to changing opinions of the electorate. The enacted plan is structured so that it will reliably elect nine Republicans and five Democrats over a wide range of studied voting patterns, with statewide Democratic vote fraction percentages ranging from the mid 40s to the low 50s, as has been typical in recent Georgia elections (see Figure \ref{histogram}). In contrast, over this same range, the non-partisan plans in our ensemble do react to the changing voter preferences by shifting the partisan make-up of those elected. Only 0.12\% (186 out of 159997) of the plans in the ensemble exhibit the same extreme non-responsiveness as the enacted plan by producing a single election outcome over the seventeen elections considered (between 2016 and 2020). Moreover, when considering the effects of modifying statewide vote fractions to start strongly favoring either party, the number of officials that would be elected for that party under the enacted plan tend to be systematically smaller than what is projected under plans from our ensemble (See Figure \ref{swing16} and Figures 8 and 9 in Supplementary Information (SI)).  
	
	\item {\bf Polarization in Competitive Districts:} We find the major cause of non-responsiveness in the enacted 2021 plan is  {\it polarization of} voters across the more competitive districts. Specifically, there are five districts that would be significantly more competitive under plans with only our non-partisan considerations, but have been shifted to become more Republican. Similarly, there are three districts that have more Democratic voters than is typical. The effect is that districts that could be more responsive to the changing opinion of the voters have become more solidified in their partisan lean and consequently the enacted plan is non-responsive to shifts in voting patterns.
	
	\item {\bf Packing in Democratic Strongholds and Cracking in a Republican Stronghold:} Utilizing a combination of our ensemble statistics and spatial analysis, we find evidence of where the most impactful changes are. We show that the heavily Democratic 4th, 5th and 13th Congressional Districts and the heavily Republican 9th District contain a significantly larger number of Democrats than typical plans in the ensemble. The consolidation or {\it packing} of Democratic voters in the 4th, 5th, and 13th Districts creates significant numbers of wasted votes and dilutes their voting power. Furthermore, based on comparisons with both the previously enacted 2011 plan and the ensemble of plans, we find that a significant number of Democratic voters have been safely added to the solidly Republican 9th District, replacing Republican voters who have been moved to the 6th and 10th Districts. This redrawing substantially weakens, or {\it cracks}, the potential influence of Democratic votes in these districts. 
\end{enumerate}

\begin{figure}
    \centering
    \includegraphics[clip=true,width=\textwidth,trim=0cm 4cm 0cm 0cm]{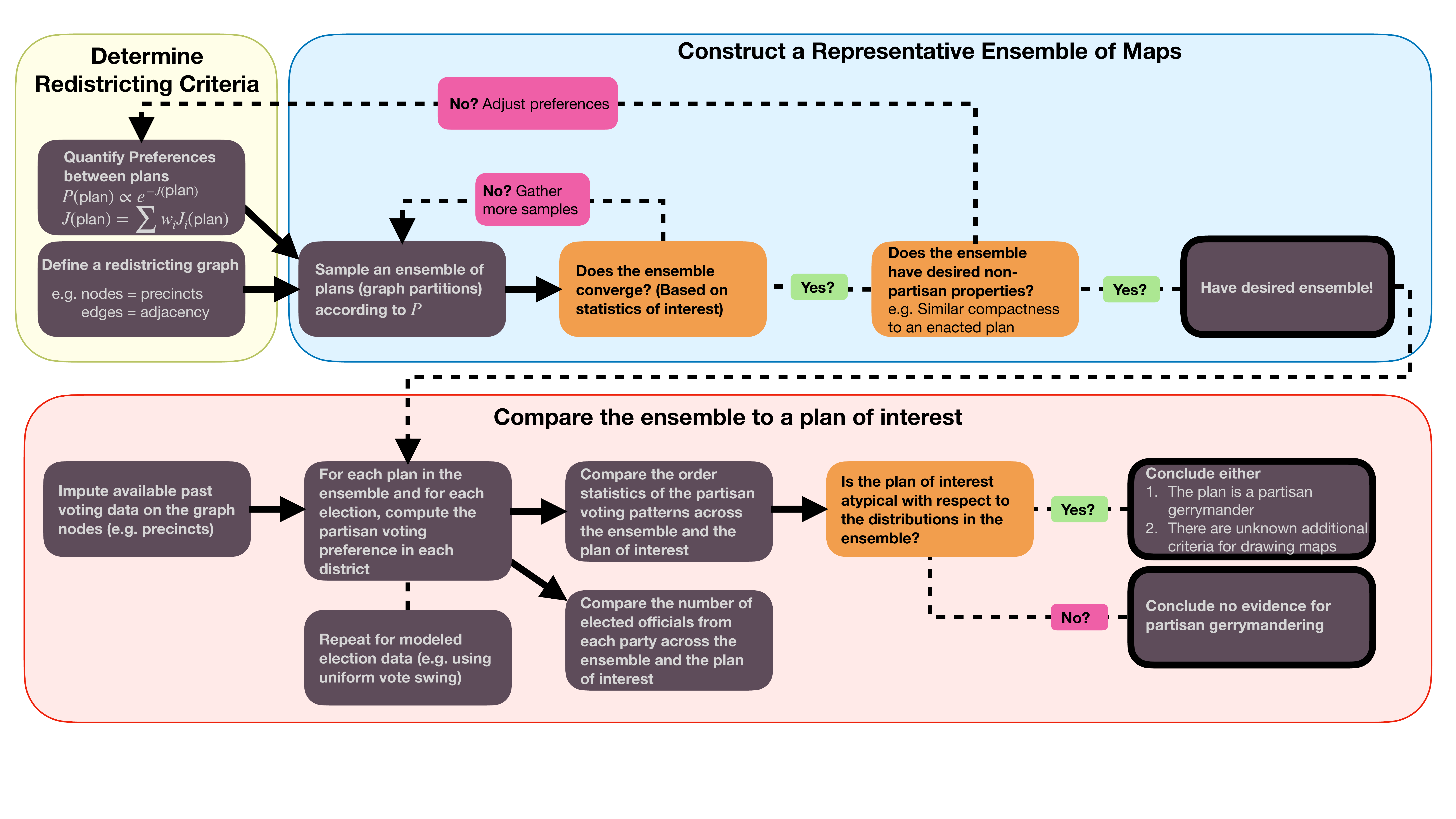}
    \caption{Outline of our procedure for detecting partisan gerrymandering. }\vspace{-0.3cm}
    \label{fig:flowchart}
\end{figure}

We summarize how the above contributions fit into a more general framework of detecting gerrymandering in a given plan, $M$, in Figure~\ref{fig:flowchart}.
This process entails three major steps: (i) designing the distribution, or family of distributions, that is used to quantify the non-partisan redistricting criteria (Section~\ref{sec:distributionfamily}); (ii) randomly sampling from the space of compliant redistricting plans according to our preferences/distribution to generate a large, non-partisan collection (or ensemble; Section~\ref{sec:howsample}); and (iii) comparing the collection of plans to a particular plan of interest, $M$ (Sections~\ref{ceah_swing}~and~\ref{why_nonresponsive}). 

The first two steps are performed iteratively, as distribution parameters are tuned so that the non-partisan criteria in the ensemble are comparable to those of a plan of interest.  When sampling from a measure, we draw samples until convergence can be established. Once a reasonable ensemble is sampled (and its convergence verified), we move to the analysis; when investigating partisan behavior of plans, we analyze both the ensemble plans and the enacted plan with past voting data. For example, we examine the distribution of elected Democrats under the ensemble coupled with a particular set of voting data; we compare this distribution with the result for the enacted plan. Of course, in this entire pipeline,  decisions must be made such as defining compactness as well as the adjacency structure and regions of the plan that may be assigned to different districts. However, the process described above is largely what has been employed in states including North Carolina~\cite{mattingly2014redistricting,nc_analysis}, Ohio~\cite{2019ohio}, Pennsylvania~\cite{mccartan2020sequential,2018league}, Virginia~\cite{deford2019redistricting}, Maryland~\cite{maryland2018}, 
and Wisconsin~\cite{herschlag2017evaluating}.
Importantly, we note that this framework does not make reference to any concept of proportionality in redistricting and elucidates how the spatial distribution of the state's electorate interacts with the redistricting process.

\section{Generating the ensemble} \label{ensemble_sec}

In this work, we focus on the Metropolized sampling methods that are capable, in theory, of sampling from any specified distribution on the space of redistricting plans.  Specifically, these include tree based approaches \cite{autry2021metropolized,autrey2021metropolizednonmulti}.  These methods also include diffusive boundary approaches, however they often run into mixing problems due to energetic barriers \cite{mattingly2014redistricting, fifield2015,https://doi.org/10.48550/arxiv.2008.07843}.  Yet other approaches rely on Metropolized methods that do not require mixing but rather compare local chain properties with a plan of interest \cite{Chikina_Frieze_Pegden_2017,chikina2019separating}.  Metropolization is not the only way to sample from a known distribution; recently a sequential Monte Carlo method was introduced that is also capable of sampling from generic measures, although it faces the same practical mixing challenges as the tree based methods \cite{mccartan2020sequential}.

Algorithm-dependent methods are another popular choice for generating redistricting ensembles. These approaches tend to be more computationally efficient. The choice of the algorithms used to sample the plans dictates the properties of generated plans (e.g., \cite{ChenRodden13,Chen15,ChoLiu16,Liu16,deford2019redistricting}). Despite being highly computationally efficient, these methods may introduce unintended and hard-to-identify biases into the collection of plans. Nonetheless, these approaches are becoming popular in practice due to the ease and efficiency in implementation.

In each state, the methods are adapted to account for the state's specific redistricting requirements.  In each of the above listed works, the measure being sampled from was either  ignored or in some way altered for the sake of computational efficiency.  To our knowledge, there have been only two state-wide investigations that have sampled from a measure constructed solely around existing policy \cite{jcmReport,jcmRebuttal,jcmHarper2021}. These works sample the state legislative districts in North Carolina by breaking the state into small sub-regions; each of the sub-regions is more manageable to sample from using standard Metropolis-Hasting based sampling methods.

One promising avenue to sample from a desired meassure is to employ parallel tempering, which samples from a range of distributions that interpolate between a measure that is efficiently sampled and a target measure.  Parallel tempering has been used previously in redistricting \cite{fifield2015,Fifield_Higgins_Imai_Tarr_2020}.  In these works, parallel tempering was coupled with diffusive boundary Markov chains which worked well for smaller graphs, but was not shown to effectively scale to larger graphs such as statewide precinct graphs.  In this work, in contrast, we use recombination techniques; these techniques can make significant changes to a redistricting plan in a single step of the Markov chain.  The challenge in using them is that it is natural to use them to focus on spanning forest measures rather than partition measures (see below and SM section B).  Our major contribution is to show how parallel tempering coupled with recombination methods can effectively refocus a measure around relevant redistricting criteria in an enacted map.

We next discuss the steps that we take to generate our ensemble of plans\footnote{Our code and data can be accessed at \url{https://git.math.duke.edu/gitlab/quantifyinggerrymandering/2020-analysis/ga-ensemble-analysis}}, starting with defining a graph, a family of probability distributions, and our accelerated sampling method.

\noindent {\bf Defining the graph.} We define a congressional redistricting plan in Georgia as a balanced graph partition of 14 elements on a graph in which nodes (roughly) represent Georgia's precincts and edges represent precincts with shared geographic boundaries. In general, redistricting processes will preserve voting precincts. However, precincts may be comprised of discontiguous regions and ensuring that these regions all belong to the same district may require a large number of neighboring precincts (and a large number of voters) to all be confined to the same districts.  In these cases, practical considerations may lead mapmakers to assign the distinct regions making up a precinct to different districts. We discuss the exceptions and modifications to this rule in Section 1 of SI. 
We work with a planar graph, which ensures there are no discontiguous districts, and resolve some of the contiguity or density issues on a case-by-case basis. Such issues are quite infrequent, and therefore, these choices have negligible impact (if any) on the analysis.

\noindent {\bf Defining a family of probability distributions.}
\label{sec:distributionfamily}
It is important to note that we are not just generating a collection of ``random redistricting plans,'' but rather sampling a \emph{distribution} on redistricting plans which encapsulates the laws and preferences for a redistricting. This makes the preferences between plans explicit, so that they can be discussed and critiqued. Additionally, one is free to use different algorithms to sample a fixed distribution. If we only describe the distribution implicitly through an algorithm, we risk introducing unforeseen biases.

The redistricting plans in our ensemble satisfy the following:

\begin{itemize}
\item {\it Contiguity:} All districts consist of one contiguous region.

\item {\it Population balance:} The total population in each district is within 1$\%$ of the ideal district population. In \cite{herschlag2017evaluating}, the authors verify that the small changes needed to make districting plans have perfectly balanced populations do not have significant impact on the partisan results of our ensembles.

\item {\it Maximum splits:} The plan splits at most 21 counties. The number of 21 is chosen in accordance with the number of county splits in the 2021 plan.\footnote{One can expect the properties of the ensemble to change if the number of allowed county splits are increased. Setting the number of splits equal to he proposed plan allows us to mimic the non-partisan properties of the plan in question.} 

\item {\it Traversing boundaries:} Districts traverse each county boundary at most once; when a district splits a county, it may not form two discontinuous regions when restricted to that county.

\item {\it Compactness:} The distribution is concentrated on more compact plans; this reflects the General Assembly's guidelines that the plans should be compact. We have tuned the distribution so that it yields plans of a similar compactness to those of the 2021 plan~\cite{ga_guidelines}. We measure compactness with the Polsby-Popper score, which is a commonly used measure of assessing compactness. See Section 5 of SI for formal definitions of the compactness and a comparison with the enacted plan (Figure 7 in SI). 
\end{itemize}

Our chosen probability distribution over redistricting plans, from which we draw plans in our ensemble, prioritizes the desired policies and complies with legal considerations. To mathematically account for compactness of plans, we use a target measure that includes the compactness score. We target these scores so that the distribution of compactness scores in our ensemble is close to that of the enacted plan from 2021. See Section 3 of SI for more details. 

\subsection{Accelerated sampling from the distribution}\label{sec:howsample}

In order to obtain random  plans from our chosen probability distribution, we run a Markov chain with Metropolized transition probabilities with parallel tempering.
In this work, we use a Metropolized tree-based sampling method. Tree-based methods have shown promise to mix\footnote{The time needed for the observables of interest to be sufficiently close to the goal distribution \cite{jerrum1996markov}.} when used to define a Markov chain that merges adjacent districts, draws a spanning tree on the merged space, and cuts the tree into two subtrees that each represent a district \cite{deford2019recombination}. 

The work of DeFord et. al \cite{deford2019recombination} has been employed as a proposal kernel and Metropolized by Autry et. al \cite{autrey2021metropolizednonmulti} and has also inspired a sequential Monte Carlo method  \cite{mccartan2020sequential}. Modified Metropolized methods have been shown to efficiently sample on the uniform measure of balanced spanning forests (i.e., each tree in the forest represents a district or partition with roughly equal population; for more details see \cite{autry2021metropolized} and Section 2 of SI). Theoretically, these methods can sample from any measure, but in practice, the chains may not mix in reasonable time. For example, in our case study for Georgia, the measure of uniform spanning forests lead to plans that were significantly less compact than the enacted plan; however, sampling from a measure with greater affinity for compactness was unable to mix even after several days using the same method on a single chain.

Parallel tempering is a class of Markov chain Monte Carlo algorithms that constructs a path of distributions interpolating between a tractable reference distribution which mixes quickly and the target distribution which we want to sample from but unable to mix. By swapping states along the path with Metropolis probability, the mixing of the target distribution may be improved \cite{syed2021parallel}. Autry et. al \cite{autrey2021metropolizednonmulti} proposed parallel tempering as a potential mechanism to successfully access new measures, however it was never directly implemented or tested in this context. In this work, we implement such a method, adopting the multi-scale approach presented in \cite{autry2021metropolized} in order to efficiently preserve counties. We begin by gathering plans sampled from the uniform distribution of spanning forests on the hierarchical structure; this measure has been shown to be efficiently sampled by these methods and we find the same in the case of Georgia. We launch four such independent chains with random initial conditions and run each chain for 10 million proposals, saving the state every 25 proposals.  We find strong agreement between the violin plots (see Figure 3 in Section 4 of SI) which means that the observable agrees in distribution and become independent of the chain's starting point. We later use the samples we get at this measure as independent and identically distributed (i.i.d.) samples for tempering runs at base measure of $\gamma = 0$,  so that instantaneous mixing can be obtained by swapping with the base measure.

Formally, the reference distribution that is uniform on hierarchical forests, which is easy to sample from is given as 
\begin{align}
    P_{F}(T) &\propto \mathbb{1}_{\mathcal{C}}(T), \nonumber\\
    P_{F}(M) &\propto \mathbb{1}_{\mathcal{C}}(M) \sum_{T_1\in T(M_1)}\dots\sum_{T_{14}\in T(M_{14})} P_{F}(T) = \mathbb{1}_{\mathcal{C}}(M) \prod_{i=1}^{14}\tau(M_i),\label{PF}
\end{align}
where $T(M_i)$ is the collection of spanning trees associated with a district $M_i$, $T_i$ is a particular spanning tree on district $M_i$, $T=(T_1,\dots,T_{14})$ is a spanning forest with trees on each of the districts in $M$, $\tau(M_i)$ is the number of hierarchical spanning trees associated with district $M_i$, and $M$ is a plan consisting of $14$ districts.  Finally, $\mathbb{1}_{\mathcal{C}}$ is the indicator function that is 1 when $M$ is in the set of constraints listed in Section~\ref{sec:distributionfamily}. Ideally, the target distribution for sampling would yield
\begin{align}
    P(M) \propto \mathbb{1}_{\mathcal{C}}(M)\exp(-w \cdot J(M)),\label{P}
\end{align}
where $J$ is a score function measuring the level of compactness in $M$, and $w$ is the weight  we can tune to match the compactness of our ensemble with the enacted plan. However, this target distribution is currently infeasible. As shown in \cite{autrey2021metropolizednonmulti}, one can interpolate between the tractable reference distribution and the intractable target distribution via a sequence of distributions parameterized by $\gamma$, given by 
\begin{align}
    P_\gamma(M)  \propto \mathbb{1}_{\mathcal{C}}(M)\prod_{i=1}^{14}\tau(M_i)^{1-\gamma} \exp(-\gamma \cdot w \cdot J(M)),\label{interP}
\end{align}
where $\gamma \in [0, 1]$. When $\gamma=0$, we sample from the uniform measure on hierarchical forests $P_F$ in~\eqref{PF}, and when $\gamma=1$ we sample from the target measure $P$ in~\eqref{P}.

We begin by naively implementing a parallel tempering scheme \cite{syed2019non} on 32 cores, setting the sequence of $\gamma_i = (i-1)/31$, where $i = 1,2,\dots, 32.$ 
Furthermore, we use the previously sampled and converged ensemble at $\gamma = 0$ as an i.i.d. sampler.  This confers the advantage that our chains instantaneously mix when they exchange with the $\gamma=0$ measure, and we later referred to it as the heat bath method.
Unfortunately, we do not observe mixing in this case. To estimate the number of needed cores, we can  draw random pairs from our samples at each level of $\gamma$ and determine the spacing of the next $\gamma$ that would be needed to ensure at least a certain percent swap probability. In a similar example, we have found that the required spacing is nearly the same for each level of $\gamma$, and that we would need somewhere between 1000 and 10000 cores to effectively implement a parallel tempering scheme, which is infeasible given our resource limits. Hence we instead make a concession to draw distributions from
\begin{align} \label{concessionP}
    P^{'}_{\gamma}(M) \propto \mathbb{1}_{\mathcal{C}}(M) \prod_{i=1}^{14}\tau(M_i) \exp(-\gamma \cdot w \cdot J(M)),
\end{align}
which favors plans with higher spanning tree counts but also has bigger probability weights for plans with our desired compactness weight $w$. In this way, our target distribution to sample from now yields
\begin{align}
    P^{'}(M)  \propto \mathbb{1}_{\mathcal{C}}(M)\prod_{i=1}^{14}\tau(M_i) \exp(- w \cdot J(M)).\label{Pnew}
\end{align}
Sampling from this new measure is computationally feasible given our available resources, and relaxes the tension between sampling from a desired distribution and the computational feasibility of sampling. With this setup, we find strong evidence of mixing using only 10 levels of $\gamma$ (i.e., 11 cores with $\gamma_{i} = (i-1)/10$, where $i = 1, \dots, 10, 11$ and the i.i.d. samples drawn at heat bath of $\gamma=0$). See Section 4 of SI for a convergence study of this method.

Although our ensemble still favors plans with higher spanning tree counts, the above scheme provides a significant advance to previous implementations in the literature.  First, it focuses on a tunable measure that cannot be readily sampled with published methods. The tempering scheme, coupled to the heat bath method at the lowest~$\gamma$, provides an efficient algorithm that can converge in an achievable amount of wall clock time with feasible computational resources. 

\vspace{-0.1cm}
\section{Non-responsiveness of the 2021 plan} \label{ceah_swing}

We see that the enacted Georgia plan does not produce different partisan results, in terms of the representative elected, over the range of recent voting patterns.  In contrast, plans in the ensemble typically respond to the changing popular vote over these same voting patterns.
Furthermore, when the Democratic vote share grows, the enacted plan systematically under-elects Democrats. There is a smaller range of election environments where the plan underelects Republicans. 

Note that a plan that reacts by changing representation when the number of votes for a particular party changes sufficiently is a minimal requirement of a democratic process responsive to the changing will of the people. To investigate this, we analyze the enacted Georgia plan by evaluating how many Democrats it would elect under a number of statewide voting patterns and compare this with the ensemble of plans. We demonstrate this through a type of plot we call \textit{collected seat histograms.} The election data we use is either a set of historical elections (Section \ref{sec:csh1}) or data generated by applying a uniform swing to a particular historical election\footnote{In accordance with the uniform swing hypothesis, we take a single election and then uniformly increase or decrease the vote percentage for a given party across all the districts. This creates a new set of voting data with the same spatial structure but a different statewide partisan percentage for each party.} (Section \ref{subsec:house_uniform_swing}). Both kinds of collected seat histograms are effective at identifying plans that are non-responsive or under-respond to changing voter opinions.

\begin{figure*}[!t]
	\centering
	\includegraphics[width=0.75\linewidth]{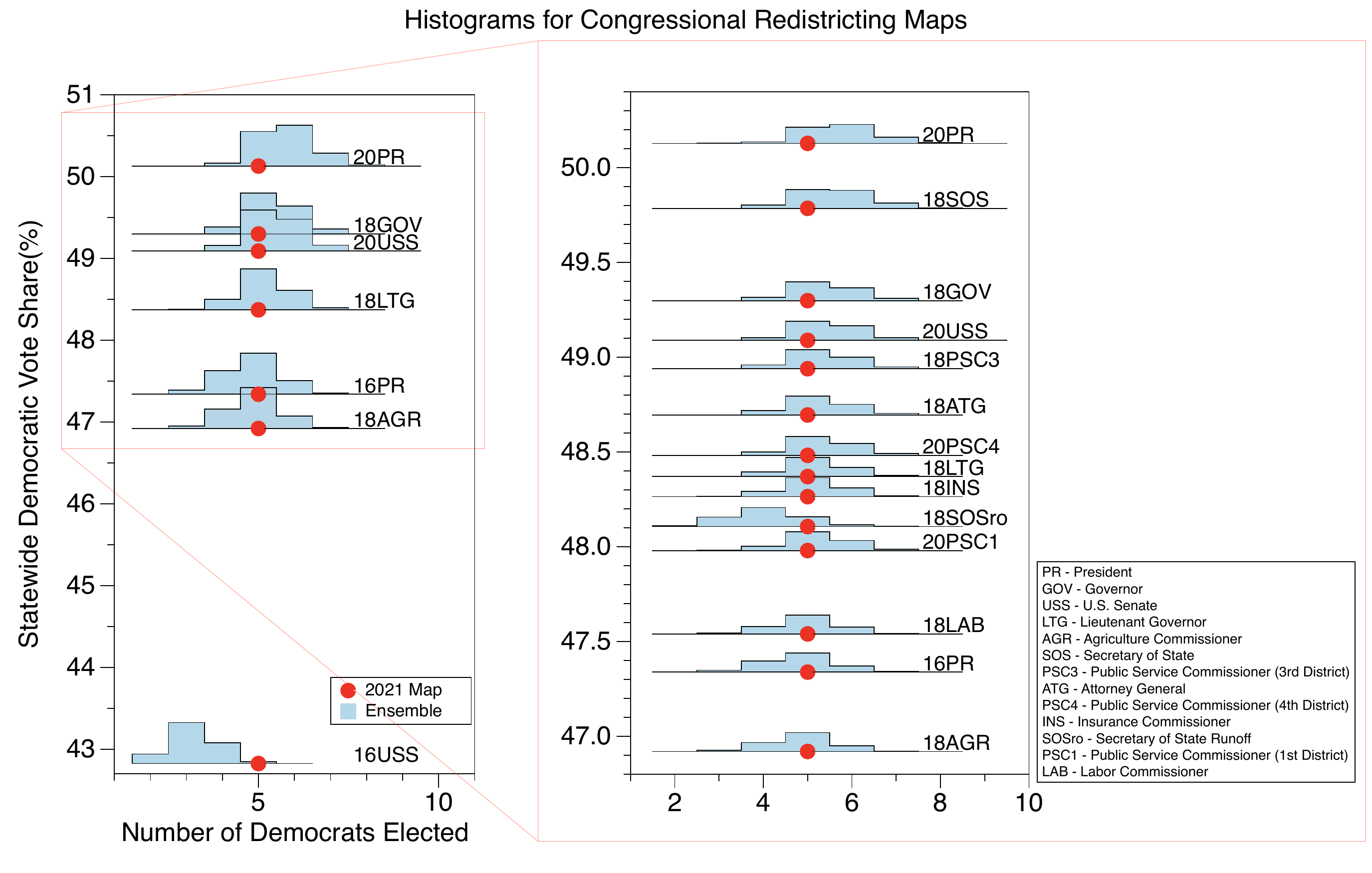}
	\caption{Each histogram represents the range and distribution of possible Democratic seats won in the ensemble of plans; the height is the relative probability of observing the result. We include only a selection of the historic vote counts for clarity so that we have a representative election for vote fractions that are spaced out. On the left axis, we provide selected Democratic statewide vote percentages. The red dots represent the reslts from the enacted 2021 plan under the various historic votes.}\vspace{-0.3cm} \label{histogram}
\end{figure*}

\subsection{Collected seat histograms under historic elections}\label{sec:csh1}

In this section, we plot a {\it collected seat histogram (CSH)} for historic elections data , i.e.,  the number of Democrats elected by the 2021 plan, and compare this against the distribution of Democrats elected across all the plans in our ensemble for historic elections, which provide different voting patterns (see Figure \ref{histogram}). These CSH plots illustrate the level of responsiveness to changes in the votes one should expect of plans drawn without a partisan bias\footnote{Throughout, we normalize election results by using the fraction
$$
\text{Democratic vote share} = \frac{\text{number of Democratic votes}}{\text{number of Democratic votes}+\text{number of Republican votes}},
$$
since the fraction of third party votes varies for each election. To determine the number of seats won by each party, we compare this fraction to~$0.5$.}. 

We see that the enacted plan elects 5 Democrats and 9 Republicans under all 17 of the historic elections we have examined.  We check whether this lack of response to voting patterns is commonly observed in the ensemble of plans.  When looking at the statewide Democratic vote share, 16 of the 17 elections are clustered around a statewide Democratic vote share of 46.5\%-50.5\%.  The one exception to this is the 2016 US Senate race which has an atypically high Republican vote share in which less than 43\% of the vote went to the Democratic candidate.
When looking at all but the the 2016 US Senate votes, only 2.1\% (3428 of the 159997) of the plans in the ensemble elect the same number of officials from each party under all 16 historic voting patterns.
If we include the US Senate 2016 votes, only 0.12\% (186 out of 159997) of the plans in the ensemble elect a fixed number of Democrats under all 17 historic voting patterns. In short, the plans in the ensemble are nearly always more responsive than the enacted plan.

It may be temping to look at Figure~\ref{histogram} and conclude that the enacted plan is fairly typical of the ensemble, as it is in the center of the histograms over the majority of the elections.  However, the plans in the ensemble elect 4-6 Democrats under the 2018 Commissioner of Agriculture election (at a statewide Democratic vote fraction of just under 47\%)  and 5-7 Democrats under 2020 Presidential election (at a statewide Democratic vote fraction of just over 50.2\%).  This shift reflects that under a typical plan it is normal for the composition of the delegation to change as the vote profile does. It is highly unusual for the composition to not change over the range of elections, as seen in the enacted plan.

In particular, the enacted congressional plan is stuck electing five Democrats in the fourteen districts, despite shifts in the statewide vote fraction and the distribution of votes across the state. Over these elections, with a statewide vote Democratic vote ranging from 42.8$\%$ to 50.1$\%$, the number of Republicans and the number of Democrats elected does not change at all. This shows the enacted plan to be highly non-responsive to the changing opinion of the electorate, and without holding the election, one largely knows that 9 Republicans and 5 Democrats will be elected.

\begin{figure}[t]%
	\centering
	\begin{subfigure}[ht]{.45\linewidth}%
		\label{4a}%
		\includegraphics[width=\linewidth]{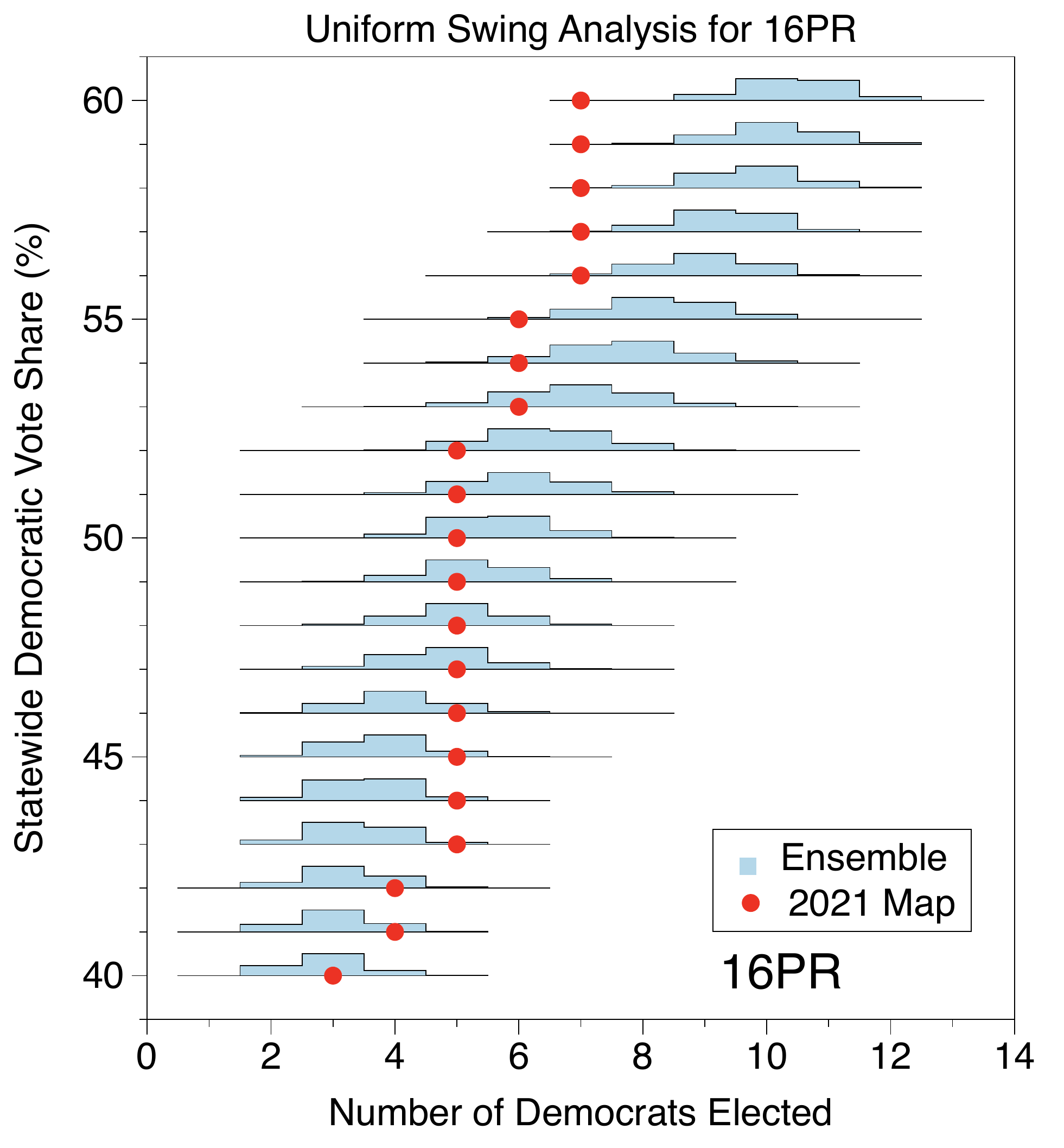}
		\end{subfigure}%
		\hspace{18pt}%
	\begin{subfigure}[ht]{.45\linewidth}%
		\label{4b}%
		\includegraphics[width=\linewidth]{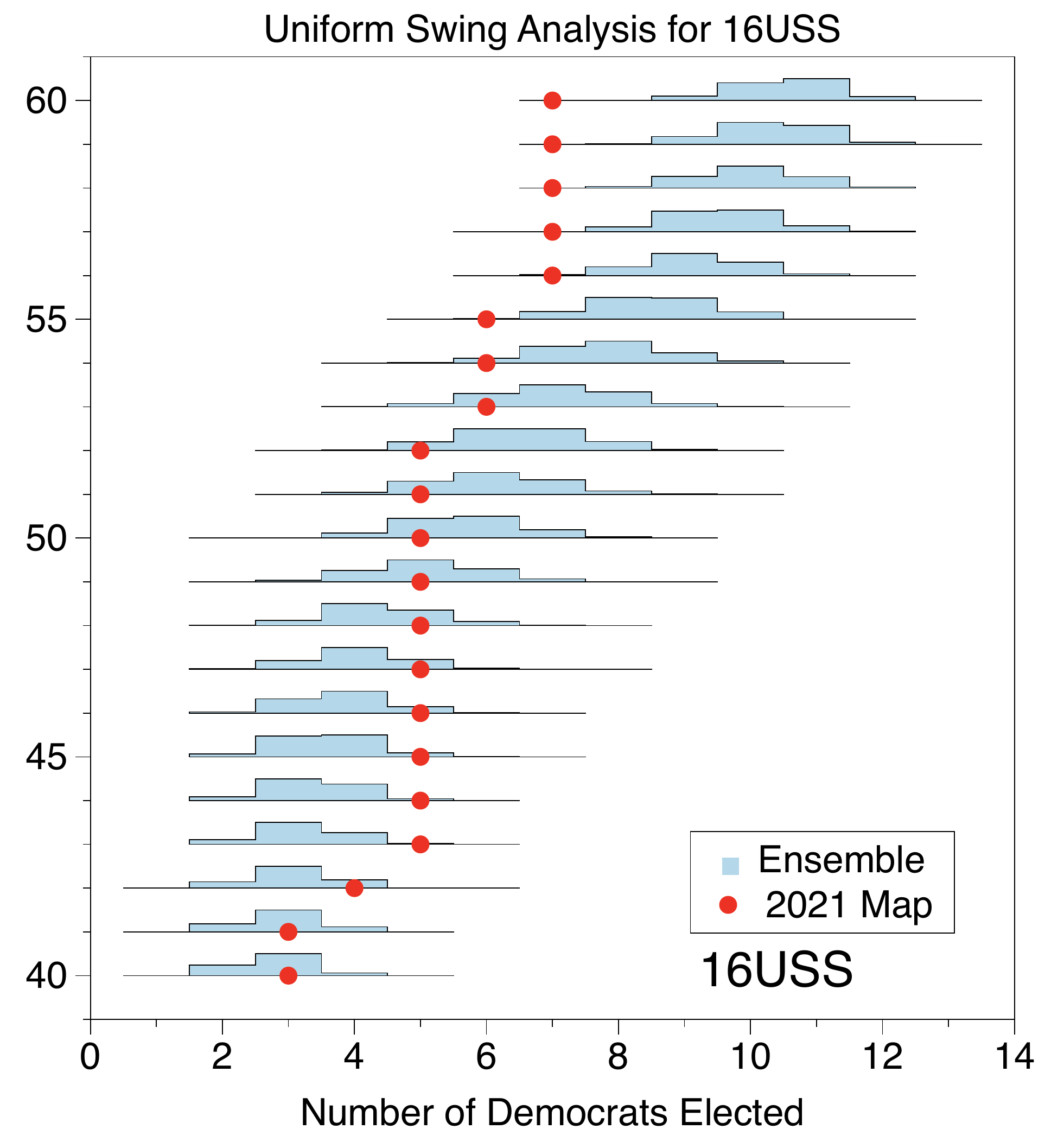}
		\end{subfigure}%
		\caption{The individual histograms give the frequency of the Democratic seat count in the ensemble for each of the shown statewide elections, with a uniform swing. The histograms are organized vertically based on the swung statewide partisan vote fraction. The red dots denote the Democratic seat count for the enacted plan for each of the swung vote profiles.}\vspace{-0.3cm}
	\label{swing16}%
\end{figure}

\subsection{Uniform swing analysis}\label{subsec:house_uniform_swing}
In addition to using historical statewide votes to produce our collected seat histograms, we create a set of collected seat histograms built from a single historical vote which is shifted in accordance with the ``uniform swing hypothesis'' to produce a new collection of votes~\cite{jackman_2014}. This preserves the relative voting pattern across the state while allowing us to study the effect of shifting the partisan tilt of the election.

In Figure~\ref{swing16} for 2016 elections (and Figures 8 and 9 of SI, for 2018 and 2020 elections respectively), we see that the non-responsiveness phenomenon from Figure~\ref{histogram} is repeated much more severely. In many cases the enacted plan fails to respond to the shifting will of the electorate, or is significantly less responsive than the plans produced by the ensemble. Our analysis shows that even if the Democratic vote share were to increase to greater than 54$\%$, the enacted plan under-elects Democrats. For example,  when the statewide Democratic vote fraction for the 2016 presidential election (16PR) and 2016 U.S. senate election (16USS) elections is swung to between 54\% and 60\%, the ensemble elects the same or fewer Democrats in 0.28\% and 0.15\% of the plans respectively. Though Georgia has not historically seen such large Democratic swings, this plan may serve to solidify these districts against future demographic changes. Furthermore, there are regions in which the enacted plan elects more Democrats then expected; for example, when the 2016 Presidential votes are swung to a Democratic statewide vote share between 43\%-45\% (see left Figure~\ref{swing16}).
This non-responsiveness is due to polarized districts that abnormally separated Democratic and Republican voters; we demonstrate this in the next section. See Section 6 of SI for additional plots for 2018 and 2020 statewide elections. Moreover, it underlines the increased polarization we see in many districts.

\section{Polarization in competitive districts} \label{why_nonresponsive}
\begin{figure}[t]%
	\centering
	\begin{subfigure}[ht]{.49\linewidth}%
		\label{8a}%
		\includegraphics[width=\linewidth]{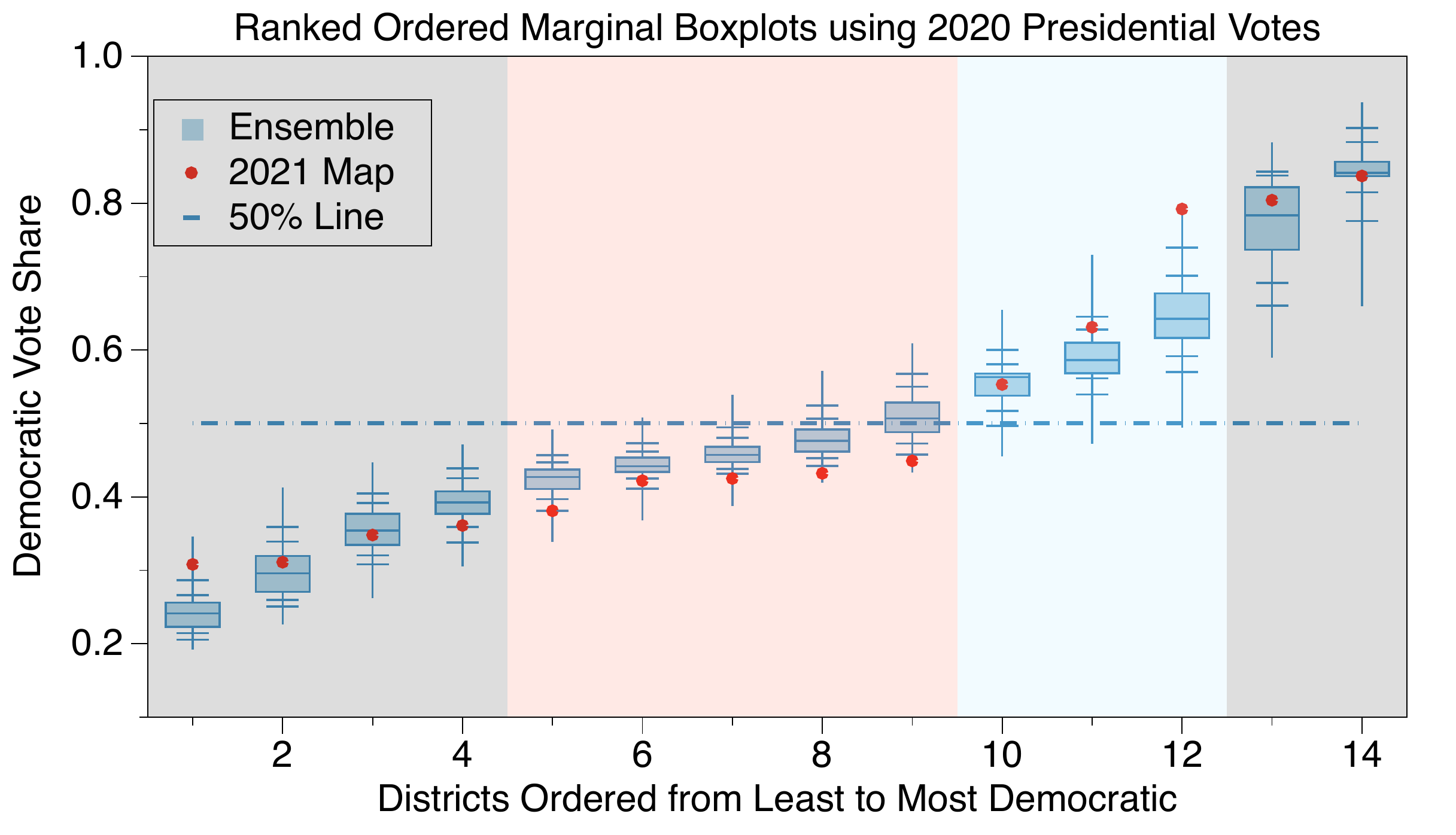}%
	\end{subfigure}
		\begin{subfigure}[ht]{.49\linewidth}%
		\label{8b}%
		\includegraphics[width=\linewidth]{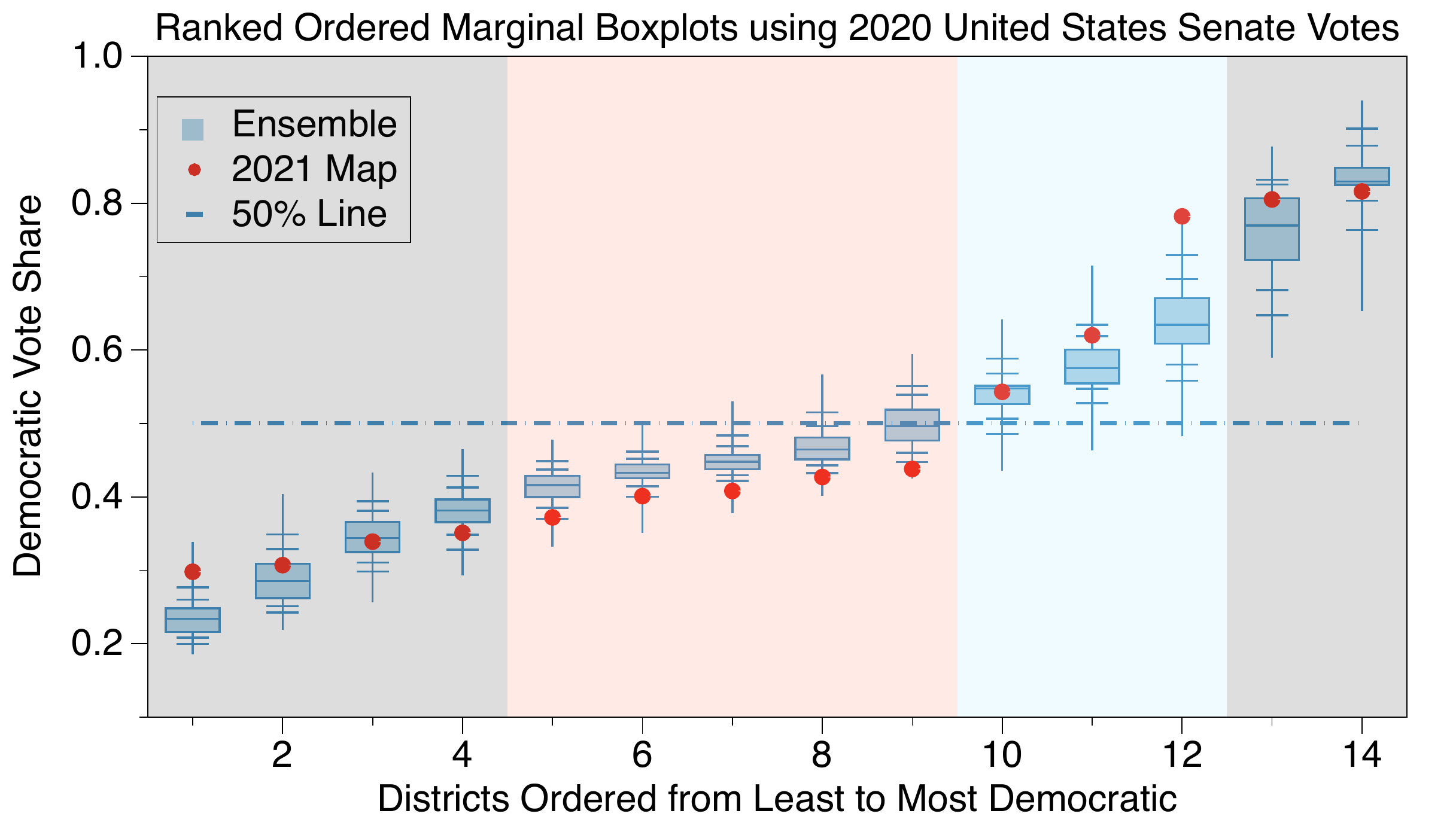}% (non-special) election.}
	\end{subfigure}
	\caption{The red dots display the ordered Democratic vote share for each of the 14 districts in the enacted 2021 plan based on 2020 presidential (top) and 2020 United States Senate (bottom) statewide elections. The box plots display the range of Democratic vote share observed for plans in the ensemble, with the 14 vote shares for each plan ranked from least to most Democratic.}\vspace{-0.3cm}
	\label{box20}%
\end{figure}

\begin{figure}[t]%
	\centering
	\begin{subfigure}[ht]{0.49\linewidth}
	    \label{5a}%
		\includegraphics[width=\linewidth]{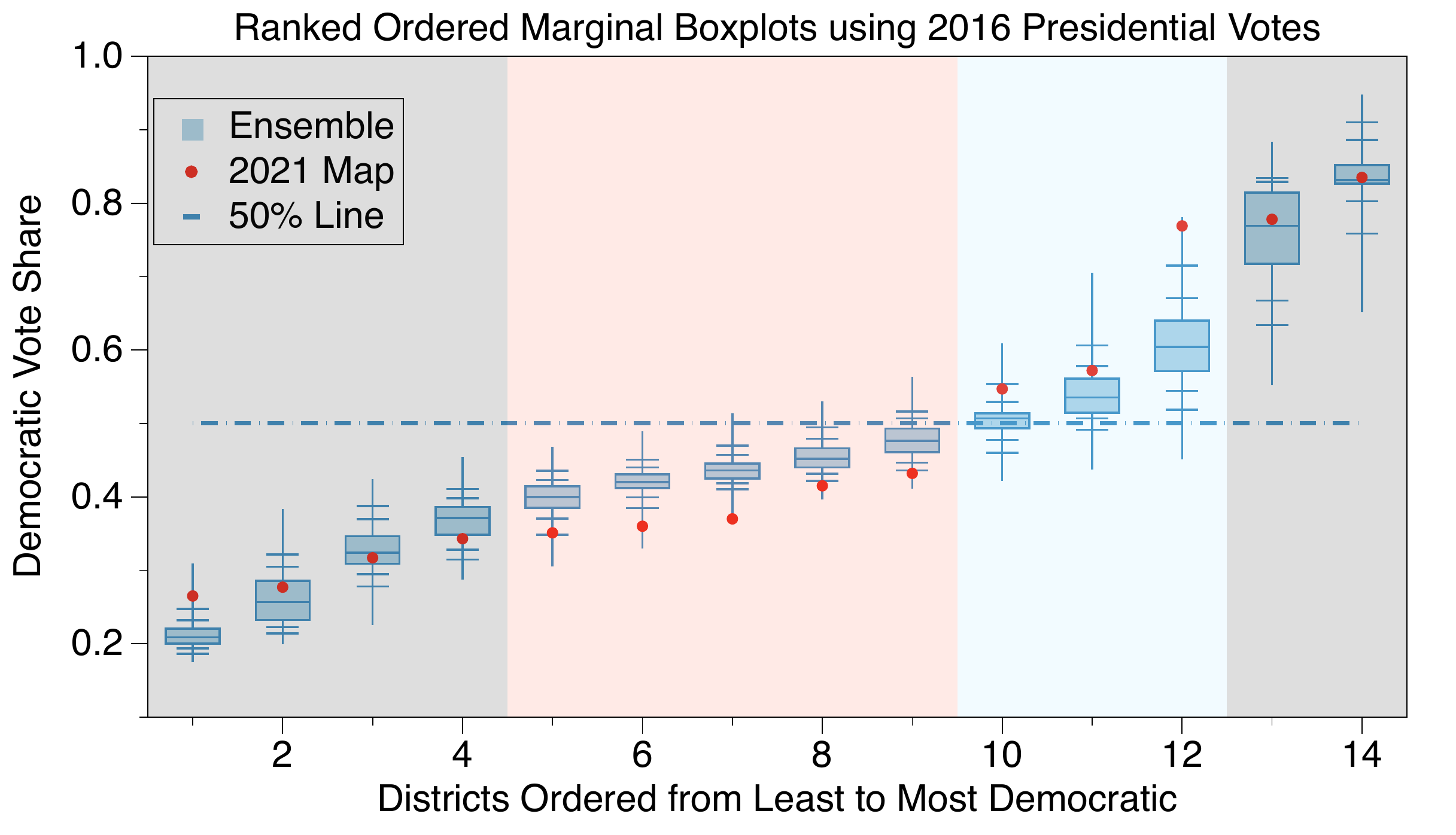}%
	\end{subfigure}
	\begin{subfigure}[ht]{0.49\linewidth}
	    \label{5b}%
		\includegraphics[width=\linewidth]{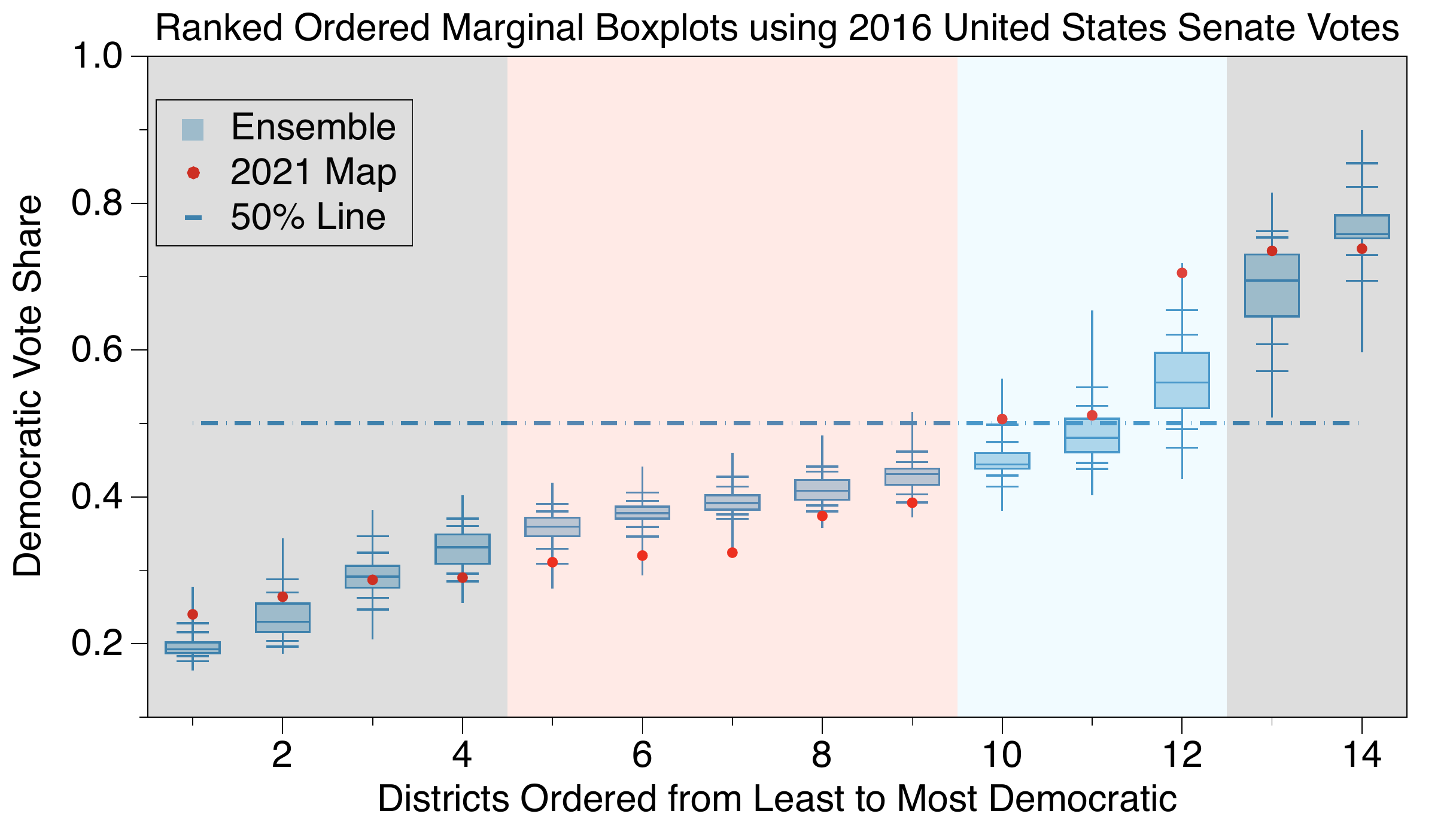}% 
	\end{subfigure}
	\caption{The red dots display the ordered Democratic vote share for each of the 14 districts in the enacted 2021 plan based on various 2016 statewide elections. The box plots display the range of Democratic vote share observed for plans in the ensemble, with the 14 vote shares for each plan ranked from least to most Democratic.
	}\vspace{-0.3cm}
	\label{box16}%
\end{figure}

In addition to looking at the number of elected representatives from each party, we examine the margins of victory within races between 2016-2020. To this end, we examine box plots that show the rank-ordered marginal distributions of the partisan vote fraction across the plans. These plots help identify when the plan contains districts with abnormally many Democrats or Republicans. This is done by considering the partisan vote fraction for one of the political parties (Democrats) in each of the districts for a given redistricting plan. These marginal vote fractions are then ordered from smallest to largest, i.e., from the most Republican district to the most Democratic district. These ordered fractions are then tabulated over all of the plans in the ensemble and used to form order statistics over the ensemble (see Figure~\ref{box20} and~\ref{box16} for 2020 and 2016 elections). Qualitatively similar results are seen for 2018 elections (Section 7 of SI).
%Appendix \ref{sec:boxplots}. 

The rank-ordered marginal box plots show the typical range of the most Republican district to the most Democratic district. Ranges are represented by box plots. In these box plots, 50$\%$ of all plans have corresponding ranked districts that lie within the box; the median is given by the line within the box; the ticks mark the 2.5$\%$, 10$\%$, 90$\%$ and 97.5$\%$ quartiles; and the extent of the lines outside of the boxes represent the range of results observed in the ensemble. Any box that lies above the 50$\%$ line on the vertical axis corresponds to a (ranked) district that will typically elect a Democrat; any box that lies below the 50$\%$ line corresponds to a (ranked) district that will typically elect a Republican (e.g., in Figure \ref{box20} (top), districts ranked 11-14 reliably elect Democrats, and districts ranked 1-6 reliably elect Republicans).

We evaluate the enacted plan with each set of votes and plot the ordered district results over the box plots. If results of particular districts lie either far above or far below the ensemble at the same ranking, this can indicate that the district was drawn to increase or decrease one party's representation within it.

In Figures~\ref{box20}-\ref{box16}, we examine a variety of elections from Figure~\ref{histogram} across 2016 to 2020 and consistently find that:
\begin{itemize}
    \item the $5^{\rm{th}}$-$9^{\rm{th}}$ most Republican districts of the enacted 2021 plan have significantly fewer Democratic votes than the corresponding $5^{\rm{th}}$-$9^{\rm{th}}$ most Republican districts of plans in the ensemble (e.g., see the pink highlighted regions),
    
    \item the $10^{\rm{th}}$-$12^{\rm{th}}$ most Republican districts of the enacted 2021 plan have significantly more Democratic votes than the corresponding $10^{\rm{th}}$-$12^{\rm{th}}$ most Republican districts of plans in the ensemble (e.g., see the blue regions). 
\end{itemize}

We consider the total Democratic votes in the $5^{\rm{th}}$-$9^{\rm{th}}$ most Republican districts from each plan in the ensemble and compare them to the total Democratic votes in the $5^{\rm{th}}$-$9^{\rm{th}}$ most Republican districts from the enacted plan. Of the 159,997 plans in our ensemble, across 17 elections from 2016 to 2020, no more than $0.017\%$ of the plans (27 out of 159,997) have the same or fewer Democratic votes than the enacted plan, which suggests that the 2021 plan polarizes voters across the $5^{\rm{th}}$-$9^{\rm{th}}$ most Republican districts to make those districts more Republican.

On the other hand, when we consider the total Democratic votes in the $10^{\rm{th}}$-$12^{\rm{th}}$ most Republican districts from each plan in the ensemble and compare them to the sum of the Democratic votes in the $10^{\rm{th}}$-$12^{\rm{th}}$ most Republican districts from the enacted plan, we observe that, across all elections, no more than $0.2\%$ of the plans (320 out of 159,997) would have the same or more Democratic votes than the enacted plan. This suggests that the 2021 plan polarizes voters across the $10^{\rm{th}}$-$12^{\rm{th}}$ most Republican districts to make those districts more Democratic. These statistics are summarized in Table 1 of SI. Consequently, districts that could have been more responsive to voters have been solidified in their partisan lean, and the enacted plan is stable and non-responsive to voting pattern shifts.

\section{A spatial analysis of packing and cracking in the 2021 plan}
In general, the ranked ordered marginal distributions do not correspond to geographic regions.
However, in Georgia, the most Republican district and the three most Democratic districts in the ensemble share geographical consistencies across most plans in the ensemble and the enacted plan. We outline the enacted districts along with a heat plan capturing typical locations of the corresponding districts in the ranked-marginals in the ensemble (see Figure~\ref{fig:geographicalSimilarity}).

The geographic locations highlighted by the heat map correspond  well to the locations of the enacted districts.\footnote{In Georgia, there is close alignment between the overarching structures of the 2011 and 2021 district plans; this is to say that the 1st District in 2011 “looks like” 1st District in 2021. We illustrate the similarity between the two plans in Figure 11 of SI.} In particular, the most Republican district in the ensemble consistently corresponds to the 9th, and the three most Democratic districts  correspond to the 4th, 5th and 13th. Using this geographic similarity, we can identify localized differences in partisan behavior of those districts in the 2021 plan and the ensemble. We find that the 4th, 5th and 13th Districts have been \textit{packed}\footnote{\textit{Packing} refers to concentrating atypically many voters of one type into a district to reduce their influence in other districts \cite{stephanopoulos2017causes}.} with Democrats, while the 9th District has been used to \textit{crack}\footnote{\textit{Cracking} is dispersing voters of one type into many districts in order to deny them a dominant voting bloc in any particular district \cite{stephanopoulos2017causes}.} the Democratic vote.

\subsection{Packing in the three most Democratic districts}
\begin{figure}
    \centering
    \includegraphics[width=0.44\textwidth]{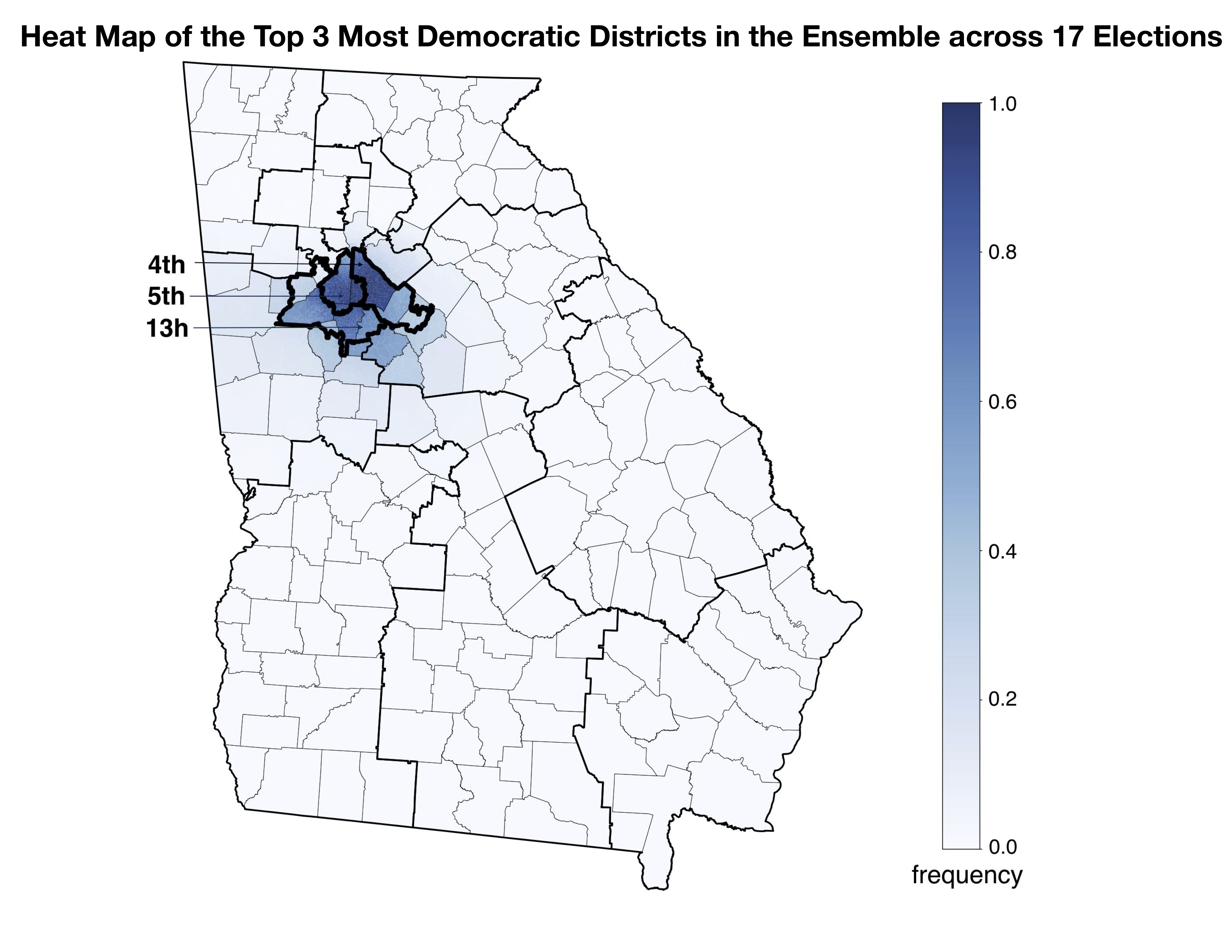}
    \includegraphics[width=0.44\textwidth]{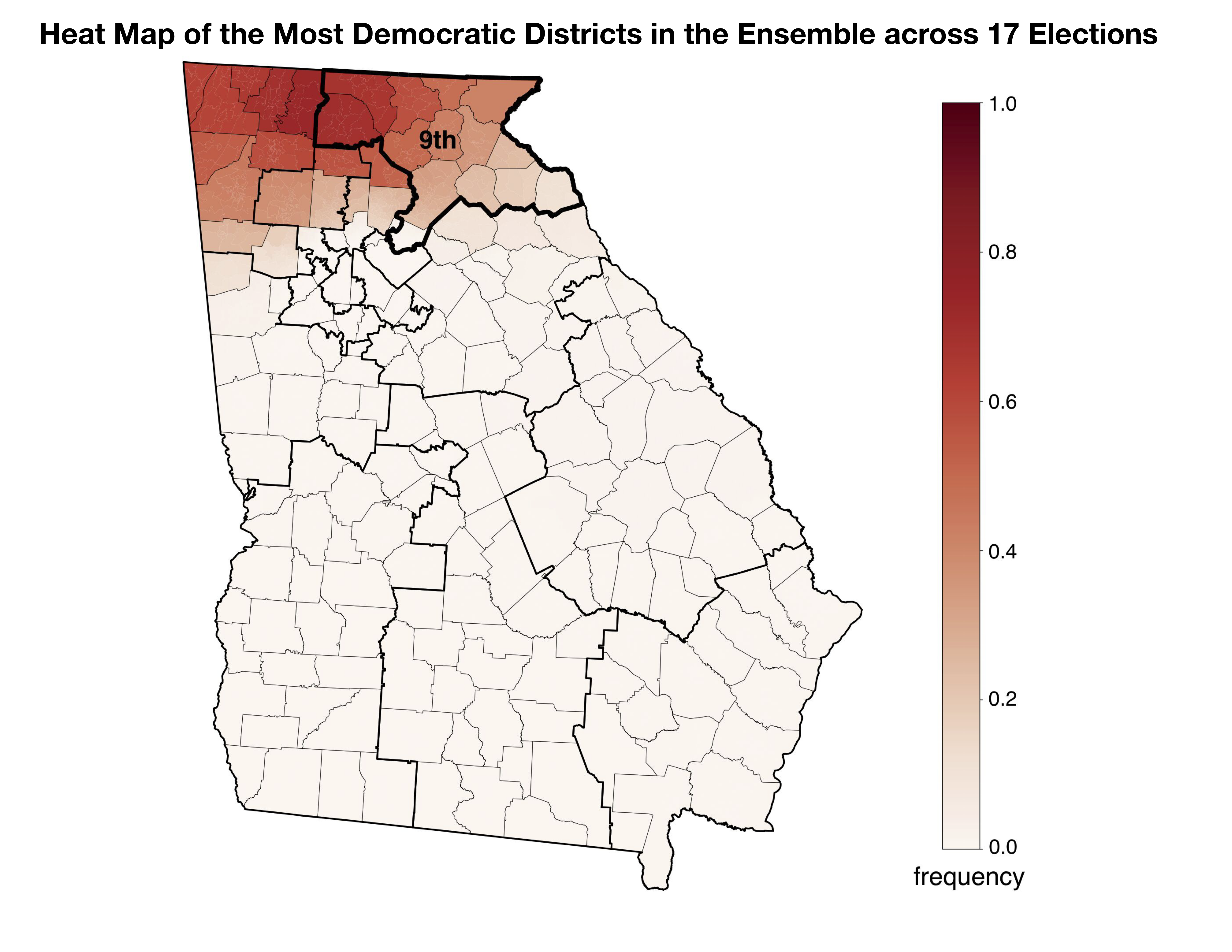}
    \caption{We display the frequency map, over the 17 statewide elections, of the proportion of plans in the ensemble in each precinct is one of the three most Democratic districts (left) and the most Republican district (right).} \vspace{-0.3cm}
    \label{fig:geographicalSimilarity}
\end{figure}

In all 17 historic elections across both the old 2011 and new 2021 enacted plans, we find that the three most Democratic districts are the 4th, 5th, and 13th Districts. 
We also examine where the three most Democratic districts occur in the ensemble by examining the frequency with which each precinct exists within the three most Democratic districts over each of the 17 elections.  We plot the resulting heat map in the ensemble and highlight the three most Democratic districts in the 2021 plan in Figure~\ref{fig:geographicalSimilarity} (left).  We find substantial similarity between the geographical location of the three most Democratic districts in the ensemble and those of the 2021 plan. 

We then compare the fraction of Democratic voters in the three most Democratic districts of the ensemble and the 2021 enacted plan.  Despite the geographical similarities, we find that the 2021 plan has the same or more Democrats than 99.41\%-99.96\% of the corresponding districts in the ensemble of plans. This demonstrates that these three districts, which correspond to a specific region of the state, have been artificially \textit{packed} with Democrats.

\subsection{Cracking in the most Republican district}
In the 2021 enacted plan, the 9th District, in the northeastern part of the state, is consistently the most Republican.\footnote{The 9th District is the most Republican across all but one (the 2020 presidential election, where it was the second most Republican district) of the 17 elections between 2016 and 2020 studied in this report.} We find that in the ensemble, the most Republican district is most often located in the northern part of the state and may either encompass precincts to the northwest or northeast (see Figure~\ref{fig:geographicalSimilarity}, right).  As above, this analysis suggests a similar geographic location of the most Republican district across plans in the ensemble and 2021 enacted plan. 

We find the most Republican district in the 2021 enacted plan contains \textbf{fewer} Republican voters than over 98.4\% of the most Republican district from plans in the ensemble across all elections. This suggests that Democrats have been atypically introduced into this District and, correspondingly, Republicans have been removed.  A direct consequence is that the newly included Democrats are removed from surrounding districts.  In our analysis below, we show that the Republican voters removed from the 9th District now dilute the voting power of Democrats in 6th and 10th Districts.

\begin{figure}
    \centering
    \includegraphics[height=4.15cm]{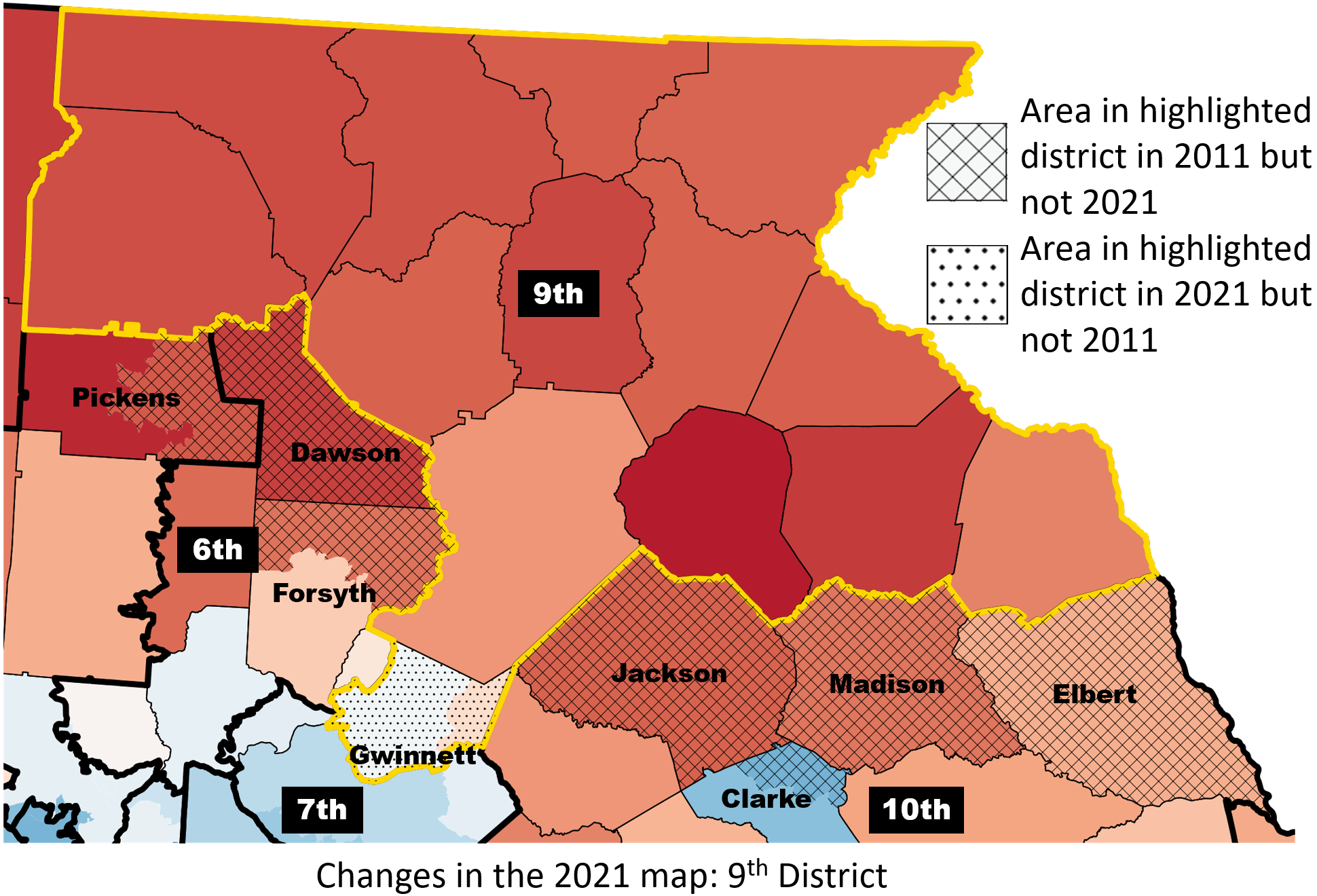}
    \includegraphics[height=4.15cm]{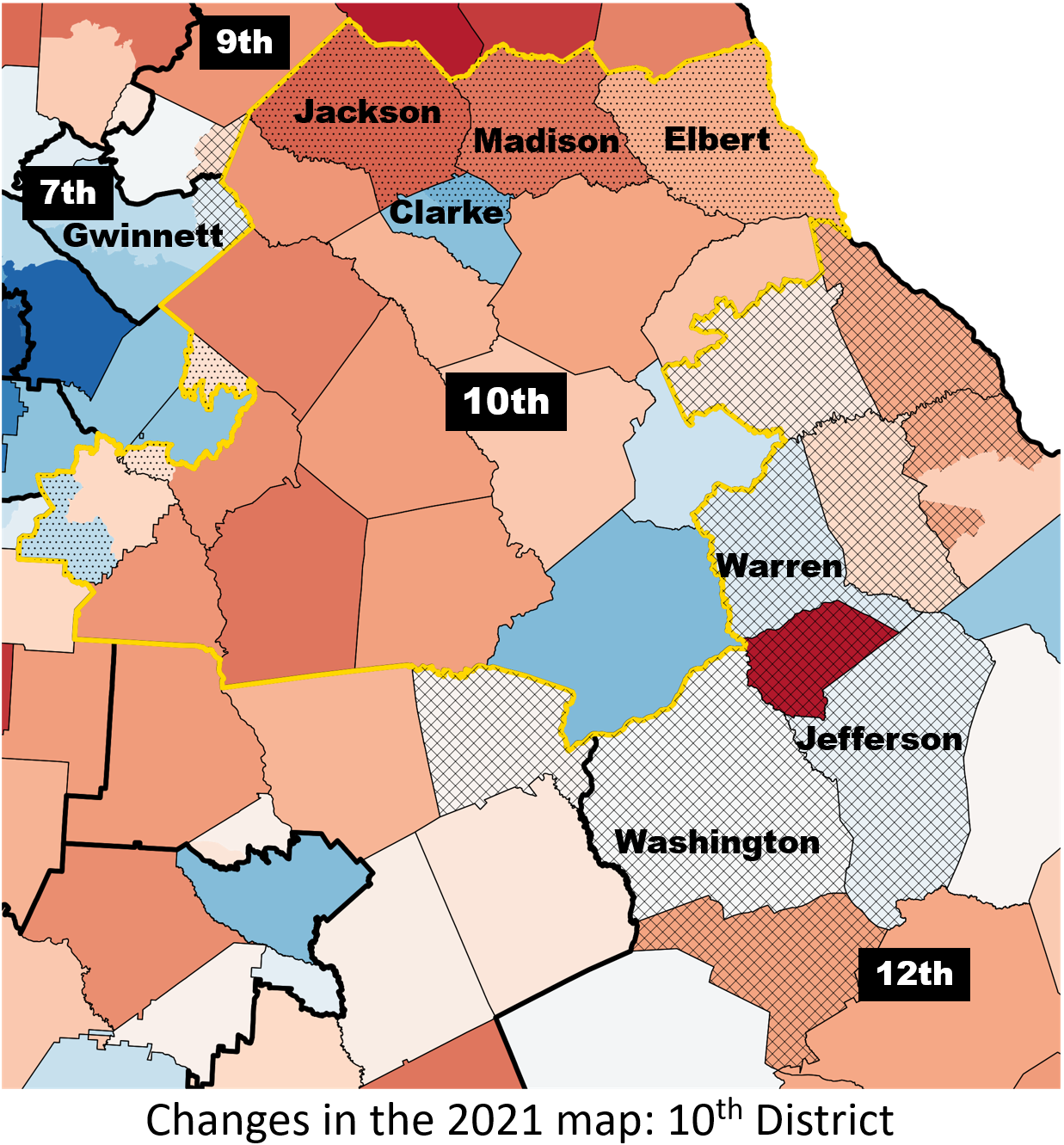}
    \includegraphics[height=4.15cm]{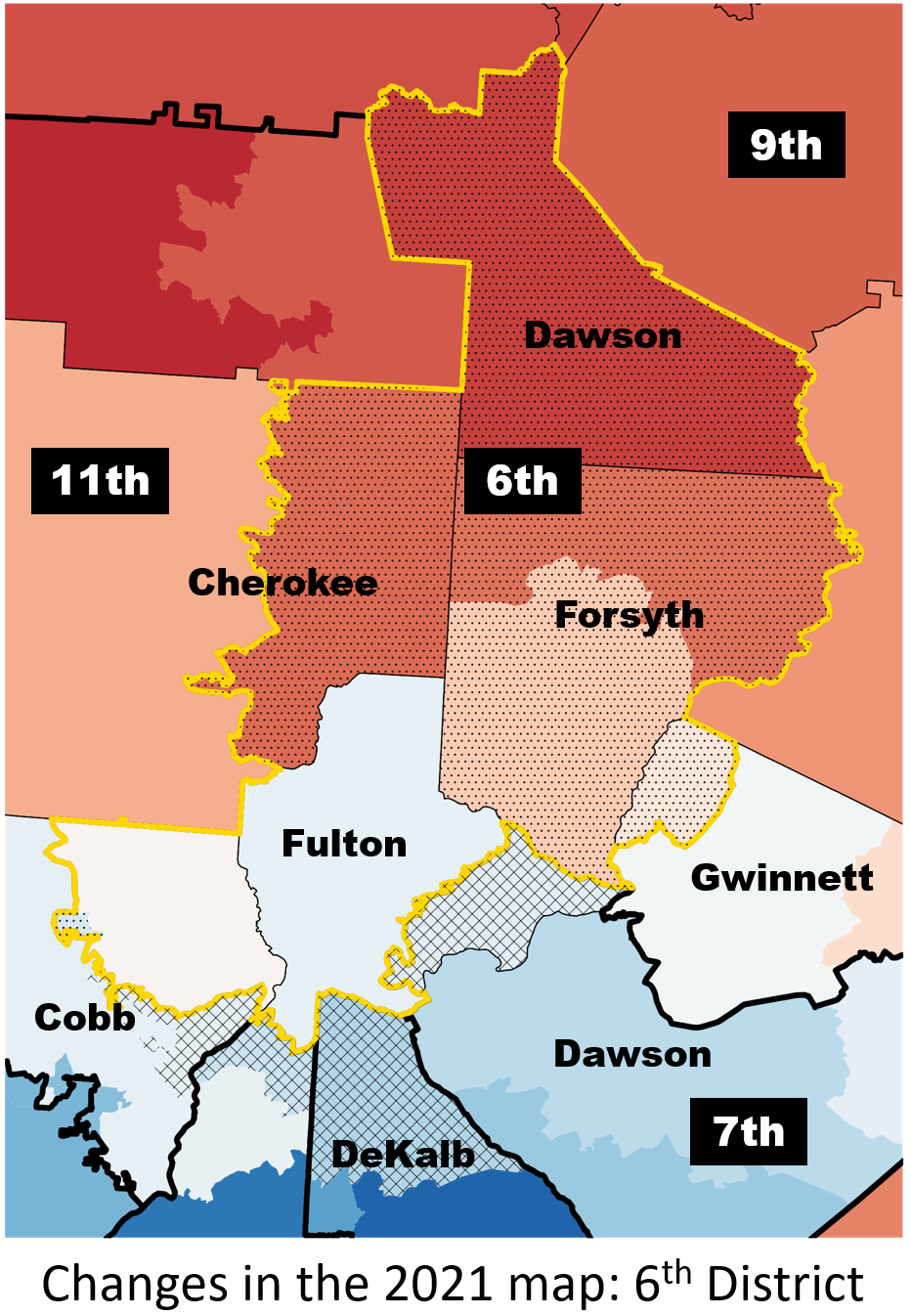}
    \caption{We display how the 2021 plan has been updated from the 2011 plan in the 9th (left), 10th (middle) and 6th (right) Districts, outlined in gold. In each plot, dots show area gained by the district from 2011 to 2021 and crosshatching shows area lost by the district from 2011 to 2021. The vote fraction of the 9th District in the 2011 plan is typical. However, in the 2021 plan, the 9th District absorbs Democratic voters (blue areas of the map) away from more competitive districts and shed Republican voters (red areas of the map) toward these districts. This move has a cascading effect that directly contributes to the changes in the 10th and 6th Districts.}\vspace{-0.3cm}
    \label{fig:shifts}
\end{figure}

To investigate \emph{where} the additional Republican voters have been moved, we begin with the following observations:
\begin{enumerate}
    \item According to the 2020 census, the 9th District as drawn in the 2011 plan had only a  1\% population deviation from being perfectly balanced in 2020. Other criteria, such as county boundaries, incumbent locations, and seats assigned to Georgia, have not changed. Therefore, this district \emph{did not} have to be substantially redrawn.
    \item The 2011 9th District is consistently the most Republican in the 2011 plan. However, its Democratic vote fraction is typical across all elections when compared to the most Republican district in each plan in the ensemble. 
\end{enumerate}
In short, the 9th District could have been almost entirely unchanged, and if so would have had a typical vote fraction in the context of the ensemble. Instead, it has been modified and is now an outlier, relative to our ensemble, with an atypically small number of Republican voters.  To determine where these Republican voters are removed, we contrast the 2011 plan with the 2021 plan in Figure~\ref{fig:shifts} (left).  We shade the regions that are no longer part of the 9th District with crosshatching and regions where the 9th District has expanded with dots.  We also color the counties (and parts of counties) based on the Democratic vote fraction in the 2020 presidential election. The 9th District has been changed to shed Republican leaning regions to the west and south, and it has expanded to the southwest to pick up Democratic voters in northern Gwinnett County.

The exchange, and consequent removal of more Republican voters from the 9th District, has a {\it cascading effect}. First, it adds Republican voters to the northern part of the 10th District.  The 10th District also recedes from the south as shown in Figure~\ref{fig:shifts} (middle). 
This motion causes the 10th District to gain Republican voters in Jackson, Madison, Elbert, and Clarke counties (once in the 9th District) and give up Democratic voters in Gwinnett County, to the 7th District, and majority-African American Warren, Washington, and Jefferson counties, to the 12th District. The 10th District is within the cluster of plans that have been depleted of Democrats, as presented in the polarization analysis in Section~\ref{why_nonresponsive}. 

Similarly, Republican voters removed from the 9th District are added to the 6th District.  This causes the 6th to move northward   picking up Republican voters in Dawson, northern Forsyth, and eastern Pickens counties (once in the 9th District) and shedding Democratic voters from part of Cobb, Fulton, DeKalb and Gwinnett counties (see Figure~\ref{fig:shifts}, right). 
These shifts dramatically increase the Republican vote fraction in the 6th District.
In the 2011 plan, the 6th District was consistently either the fourth- or fifth-most Democratic district; in the 2021 plan it is now one of the districts with atypically few Democrats and would stably elect a Republican representative, according to historic voting trends.  
Localized analysis of gerrymandering is still a developing field. A new tool has recently emerged to match districts spatially between the ensemble and enacted plans~\cite{needham2021geometric}.  We plan to utilize this tool in a follow-up study.

\section{Conclusion}
This report shows that the 2021 enacted congressional district plan in Georgia is likely to be highly non-responsive to the changing opinions of the electorate. Moreover, there is mathematical evidence of polarization of competitive districts, which has been caused in part by the redrawing of the 6th, 9th, and 10th Districts. The non-responsiveness of Georgia's congressional plan is highly improbable, even when considering effects of the Voting Rights Act in ensuring 4 of 13  districts can elect an  African-American representative (i.e., a proportional number).  We note we have not yet tested the effect of enforcing 4 (near) majority-minority districts, so it is possible that this enforcement is what lead to the lack of responsiveness.  As found in the Supreme Court Ruling in \emph{Cooper v. Harris}, it is questionable whether such an extreme concentration is legal, so we have omitted such an analysis. See more details in Section 9 of SI.

We implemented tempering techniques on existing multi-scale tree-based methods. To the best of our knowledge, this is the first time tempering has been explicitly used to sample with these techniques. These modifications have allowed us to sample from measures that were previously inaccessible. Specifically, they have allowed us to tune the Polsby-Popper compactness to match that of the enacted plan.  We remark that our ensemble \emph{is} still weighted toward plans with higher numbers of associated spanning trees and that further work is still needed to overcome this limitation; however, to our knowledge, this work has successfully sampled from the most flexible target measure on a complicated redistricting problem. Furthermore, these methods should be fully portable to other states. Although beyond the scope of the current work, we have also used these techniques to analyze Georgia's General Assembly plans.  We save a full analysis of these plans for future work, but note that these plans are also significantly less responsive than the ensemble of plans.  We display some of our results in Section 10 of SI.

\bibliographystyle{ACM-Reference-Format}
\bibliography{bib}

\pagebreak
\begin{center}
\textbf{\large Supplemental Information}
\end{center}
%%%%%%%%%% Merge with supplemental materials %%%%%%%%%%
%%%%%%%%%% Prefix a "S" to all equations, figures, tables and reset the counter %%%%%%%%%%
\setcounter{equation}{0}
\setcounter{figure}{0}
\setcounter{table}{0}
\setcounter{page}{1}
\makeatletter
\renewcommand{\theequation}{S\arabic{equation}}
\renewcommand{\thefigure}{S\arabic{figure}}
\renewcommand{\bibnumfmt}[1]{[S#1]}
\renewcommand{\citenumfont}[1]{S#1}

\setcounter{section}{0}

\section{Clarifications on the redistricting graph} \label{asec:graph}
As a rule of thumb, a redistricting plan in the ensemble is defined to be a graph partition where the nodes of the graph are represented by precincts and the edges by precincts with shared perimeters (using rook adjacency). However, some precincts are multi-polygonal; if we treat these precincts as a single node, then the graph is no longer planar and it is possible to draw discontinuous districts. Precincts are also supposed to be kept intact, where practicable, in redistricting.  Therefore, the majority of the multi-polygonal precincts are merged with their nearest neighbors to create nodes that are collections of precincts. Neighbors are selected to minimize the accumulated population in the collection of precincts. 

\begin{figure}[ht]
\centering
		\begin{subfigure}[ht]{\linewidth}
			\centering
% 			\label{7c}%
		\includegraphics[width=0.3\linewidth]{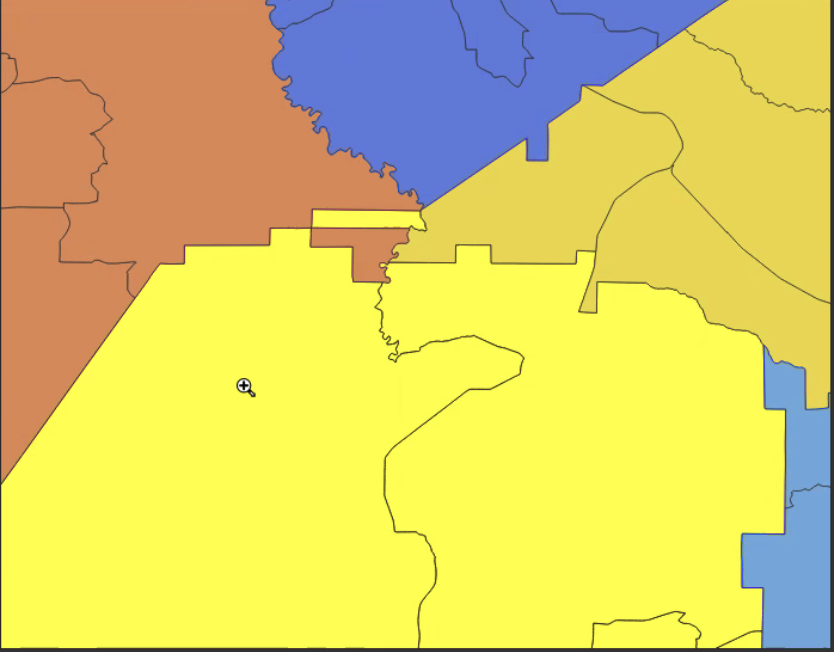}%
	\end{subfigure}
	    \caption{One of the precincts in yellow is a multi-polygon, with one small part disconnected from its county in yellow, and is in contact with two other counties. Furthermore, the orange central precinct is also multi-polygonal.  We resolve this by absorbing the small yellow northern rectangle with the surrounding orange precinct.}
	\label{GAmulti}%
\end{figure}

There are two exceptions to merging on multi-polygonal precincts.  The first is that when one of the polygons is isolated in a separate county, we do not merge, as there are examples in GA redistricting history where districts may be discontinuous in cases where counties and multi-polygonal precincts are preserved. There is one exception to this, in which a multi-polygonal precinct has a region which is entirely disconnected from its county and is in contact with two other counties in a way that would break the planar nature of the graph, as shown in Figure~\ref{GAmulti}.  In this case, we merge the extraneous region with the precinct in the other county that shares the majority of its perimeter.  Second, the above mentioned merging process generates some collections of precincts that have extremely large populations due to urban and highly complex precinct structures.  If a collected precinct has total population more than 20,000, we allow each multi-polygonal component to be its own node.\footnote{In contrast, an ideal congressional district has a population of roughly 765,136 people} In this case, it is possible that our redistricting plans in the ensemble to split these precincts into different districts, however we allow this as part of our sampling procedure.

\section{Sampling on Spanning Forests}\label{sec:app_trees}
\begin{figure}
    \centering
    \includegraphics[width=0.5\linewidth]{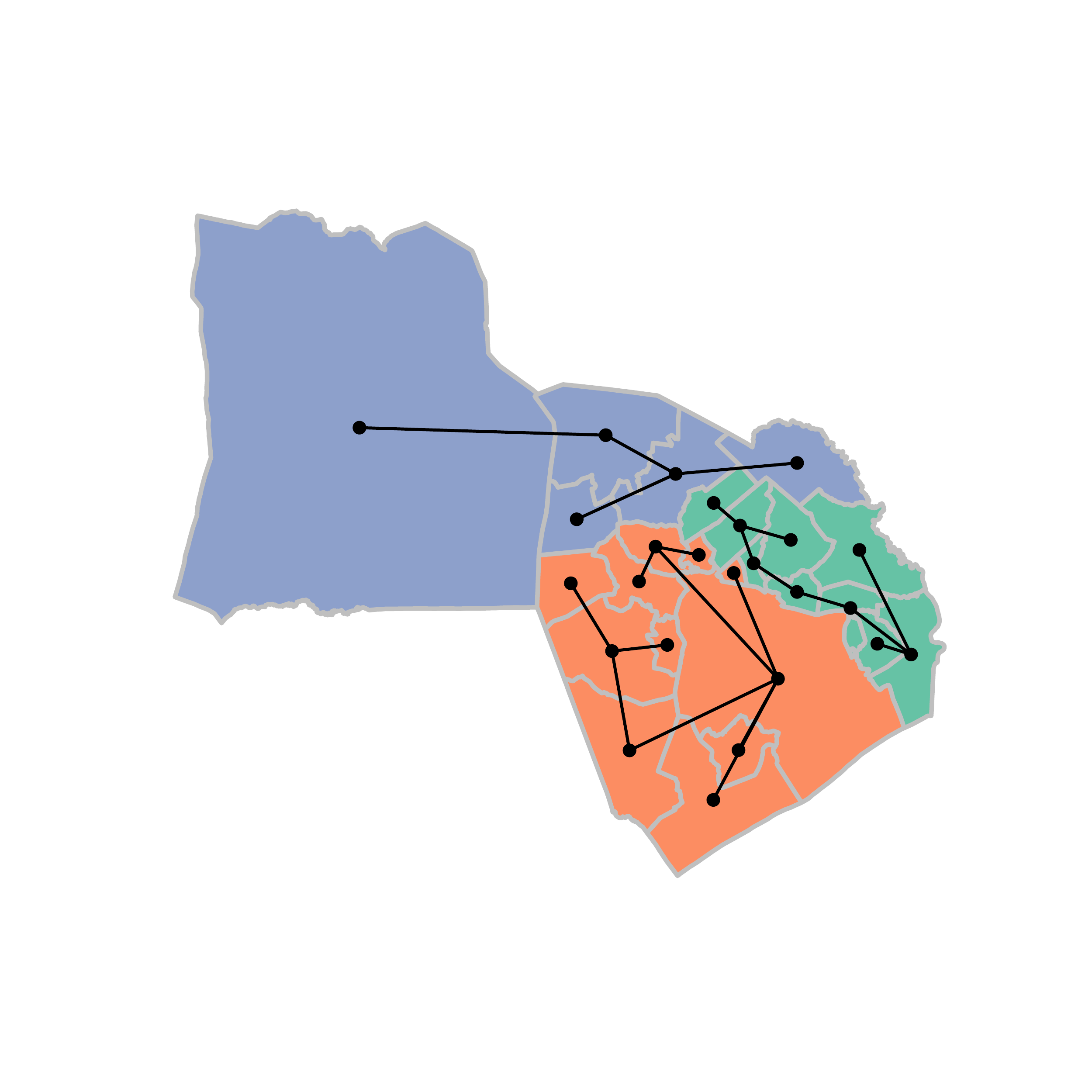}
    \caption{We show an example state with 3 districts.  Each district has an associated spanning tree (i.e. there is a spanning forest with 3 trees on the graph that represents 3 districts).  Note that we could keep the same districts, but redraw the tree in any of the districts; when sampling on uniform spanning trees, plans with more associated spanning forests are more likely to be drawn.}
    \label{fig:spanning_forest_example}
\end{figure}

The current state-of-the-art sampling algorithms rely on merging adjacent districts, drawing a spanning tree on the merged space, and then removing an edge to determine two new districts \cite{deford2019recombination, autrey2021metropolizednonmulti, autry2021metropolized, mccartan2020sequential}.  The cut spanning tree leaves two new trees, each of which represents a district.  It is then natural to associate a redistricting plan with a specified spanning forest that results from the previously cut tree on the joined space (e.g. see Figure~\ref{fig:spanning_forest_example}).

Indeed, when Metropolizing these algorithms, it is most efficient to sample uniformly on all (population balanced) spanning forests, which is to say that any spanning forest associated with a (population balanced) redistricting plan is equally likely (e.g. see \cite{autrey2021metropolizednonmulti}).  The consequence of this choice is that redistricting plans with more possible ways of drawing associated spanning forests become more likely.  

When modifying the target measure away from one that is uniform on balanced spanning forests, the proposal measure quickly becomes singular with respect to the target measure. Additional tools, such as parallel tempering (and the modifications we make to parallel tempering) are needed to alleviate the mismatch between these measures.  Below we describe how we have modified the target measure and tuned it to be more reflective of the properties of the enacted plan.

\section{The Target Measure}\label{sec:targetmeasure}
We run the sampling method summarized in the main report for up to four million steps under different random seeds. We use parallel tempering to get an appropriate  compactness weight with eleven replicas with compactness weight ranging from 0.00 to 0.04.\footnote{In tempering, we use the hyper-parameter $\gamma=0$ from the Metropolized Forest RECOM algorithm; see \cite{autrey2021metropolizednonmulti} for details.} The compactness weight enters the probability distribution on plans via an exponential distribution on the compactness score (i.e.\ the sum of the isoperimetric ratios for each district).  The distribution is then given as $\pi(T)=p(T)/\mathcal{Z}$, where $T$ is a spanning forest that defines a redistricting plan, $\mathcal{Z}$ is a normalization constant, and 
\begin{align*}
p(T) = \begin{cases}
e^{-{w^{'} J(T)}} & \text{if $T$ meets constraining criteria,}\\
0 & \text{o.w.}\end{cases}
\end{align*}
and $J$ is the sum of the isoperimetric ratios of the districts and $w^{'}$ is the compactness weight which varies from 0 to 0.04.   The isoperimetric ratio is the square of its perimeter divided by its area.  See \cite{autry2021metropolized, autrey2021metropolizednonmulti} for more details.  In our tempering framework, $w^{'} = \gamma \cdot w$. The compactness weight of $w^{'}=0.04$ is the "target" distribution which is estimated to yield districts with comparable compactness levels to the enacted plan before generating the ensemble of plans.

We can sum over all forests that define a partition and arrive at a measure on a partition, $M$, given by 
\begin{align*}
p(M) = \begin{cases}
e^{-w J(M)}\prod_{i=1}^{14}\tau(M_i) & \text{if $M$ meets constraining criteria,}\\
0 & \text{o.w.}\end{cases}
\end{align*}
where $\tau$ is the number of trees that can be drawn on a sub-graph and $M_i$ are the 14 sub-graphs or partitions that define the districts.
	
\section{Convergence}\label{convergence_app}
\begin{figure}[ht]
\centering
		\begin{subfigure}[ht]{\linewidth}
			\centering
% 			\label{7c}%
		\includegraphics[width=0.7\linewidth]{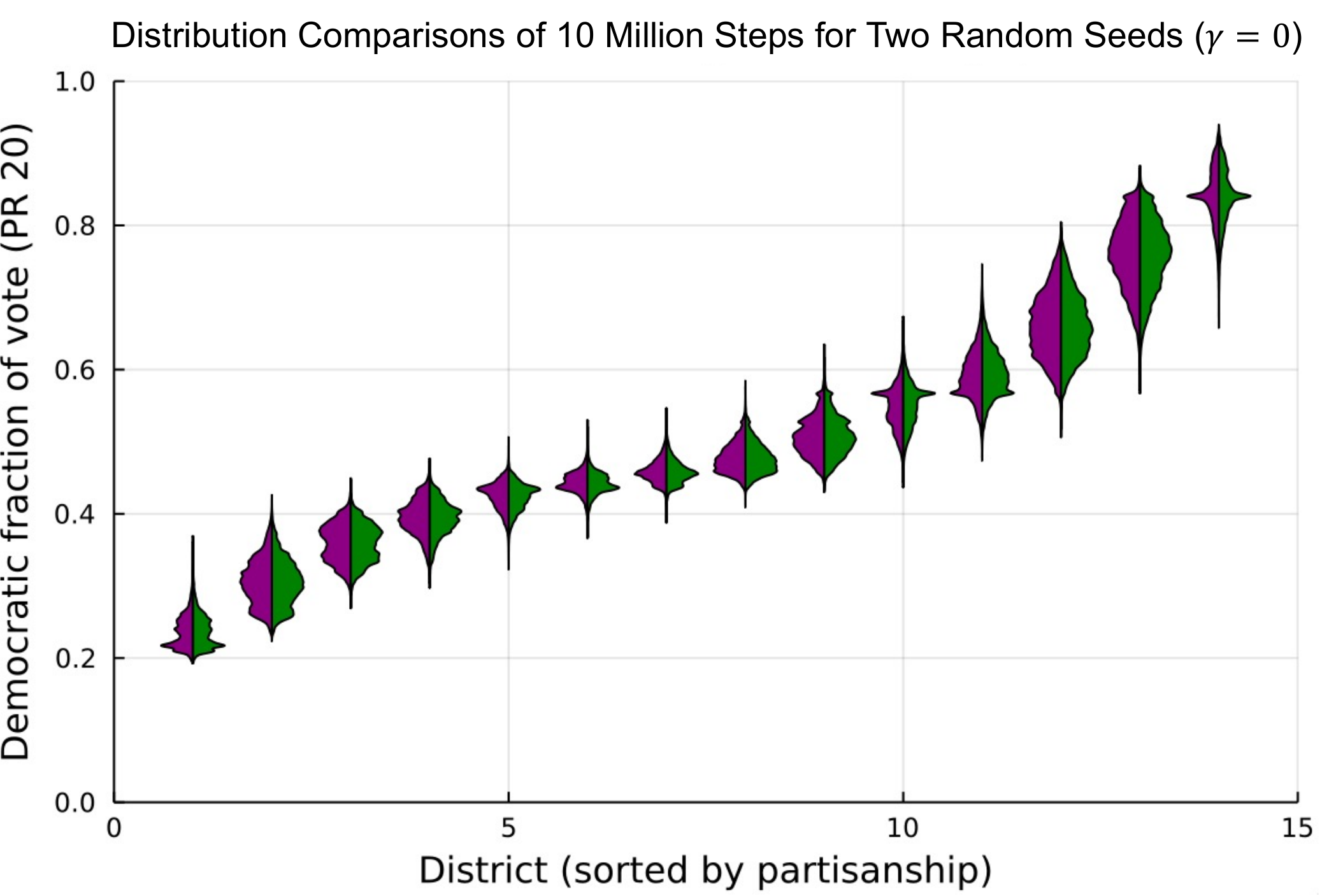}%
	\end{subfigure}
	    \caption{The violin plots show the distribution comparisons between two 10-million runs with different random seeds at the base measure of $\gamma = 0$. For each run, samples are drawn every 25 steps. }
	\label{gamma0}%
\end{figure}

We first run the sampling method with the measure uniform on the number of hierarchical spanning forests in Equation (1) of the paper to draw us i.i.d. samples at base measure of $\gamma = 0$, also known as heat bath method as defined in the main body. In order to confirm the chain under this measure is mixing and the samples are i.i.d, we launch four such independent chains, each with a different random initial condition, and run each chain for 10 million proposals (steps) and take samples every 25 steps. Figure~\ref{gamma0} shows the distribution results of random two chains out of the four, where we take the Democratic vote fraction in 20PR as an example to demonstrate the district distributions in each plan. Note we order districts in each plan from the least to the most Democratic. The great similarities of the distributions in purple and green in Figure~\ref{gamma0} shows the mixing of the chain and the samples can be taken as i.i.d. samples.

\begin{figure}[ht]
\centering
		\begin{subfigure}[ht]{\linewidth}
			\centering
% 			\label{7c}%
		\includegraphics[width=0.8\linewidth]{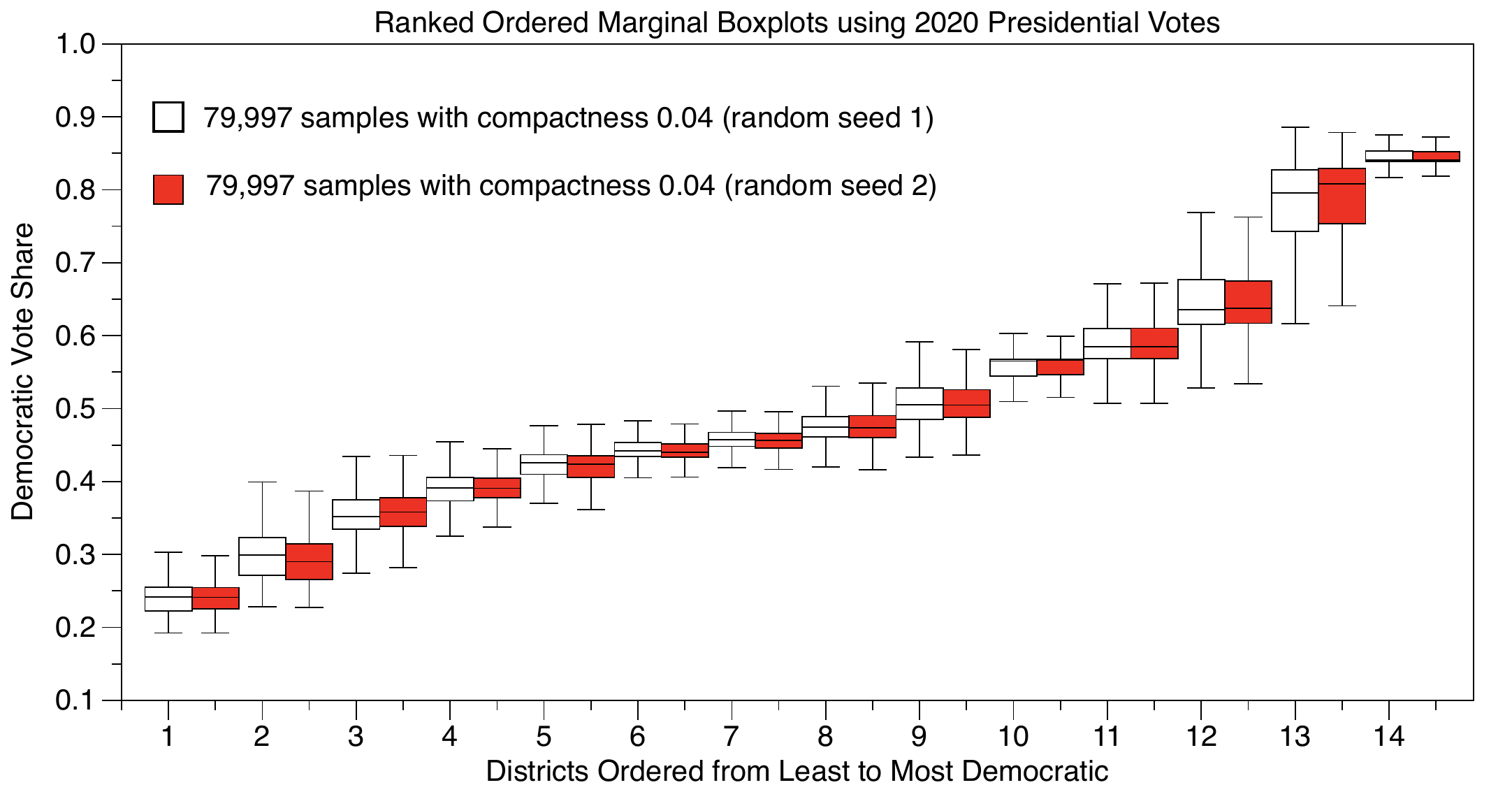}%
	\end{subfigure}
	    \caption{Democratic vote share in each district of each plan across two different sets of 79,997 plans. The two sets of plans are taken from two runs of two million steps each, starting from two different initial random seeds, demonstrating that choice of initial seed was not significant.}
	\label{convergence1}%
\end{figure}

\begin{figure}[H]
\centering
		\begin{subfigure}[ht]{\linewidth}
			\centering
% 			\label{7d}%
		\includegraphics[width=0.8\linewidth]{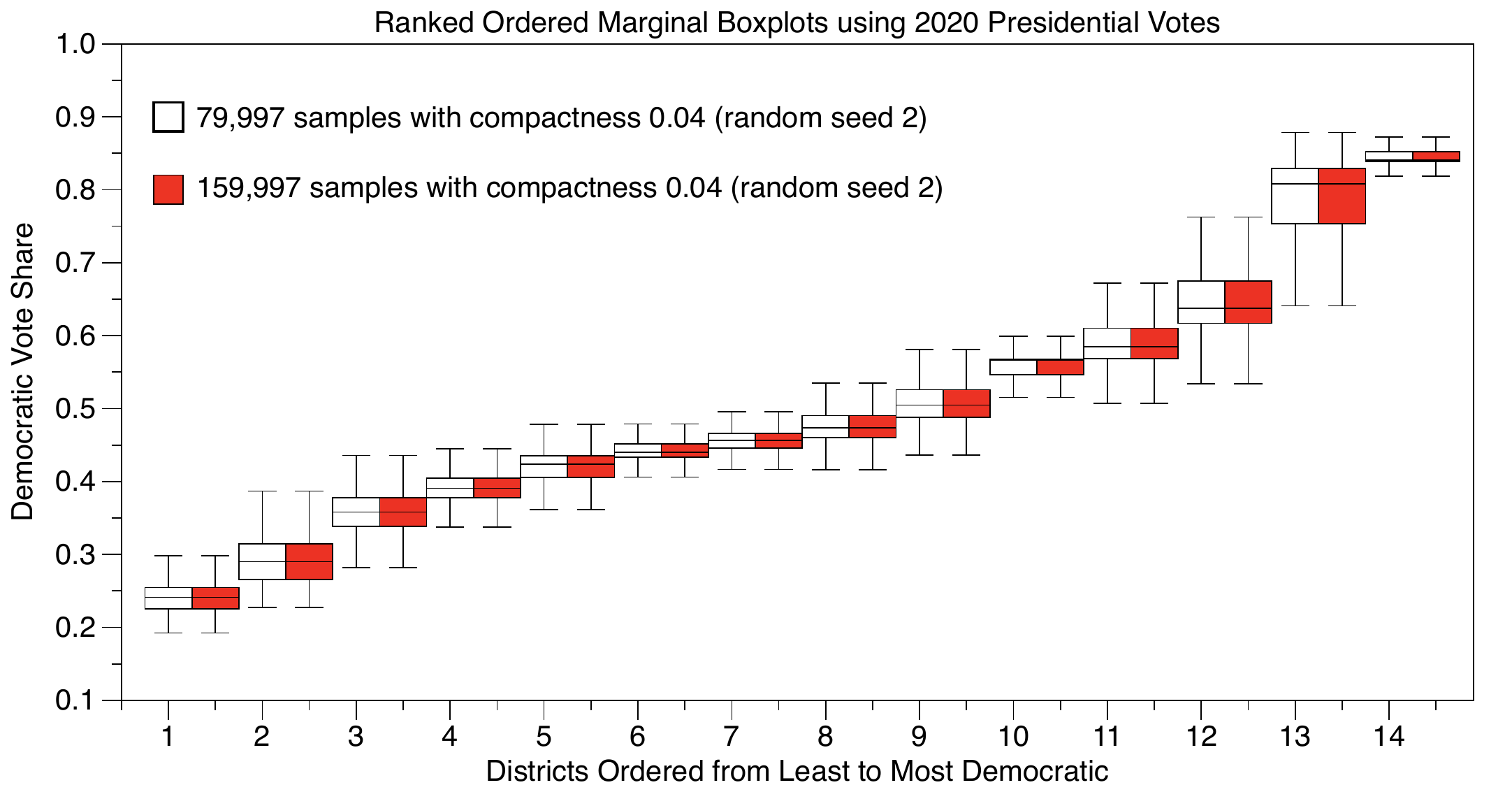}%
	\end{subfigure}
	    \caption{Comparison of the ranked 2020 presidential Democratic vote share in each district of each plan across two different sets of plans. The two sets of plans are taken from two runs of two million steps each, starting from two different initial random seeds, demonstrating that two million steps was sufficient for the run to converge.}
	\label{convergence3}%
\end{figure}

\begin{figure}[ht]
\centering
		\begin{subfigure}[ht]{\linewidth}
			\centering
			\label{7d}%
		\includegraphics[width=0.8\linewidth]{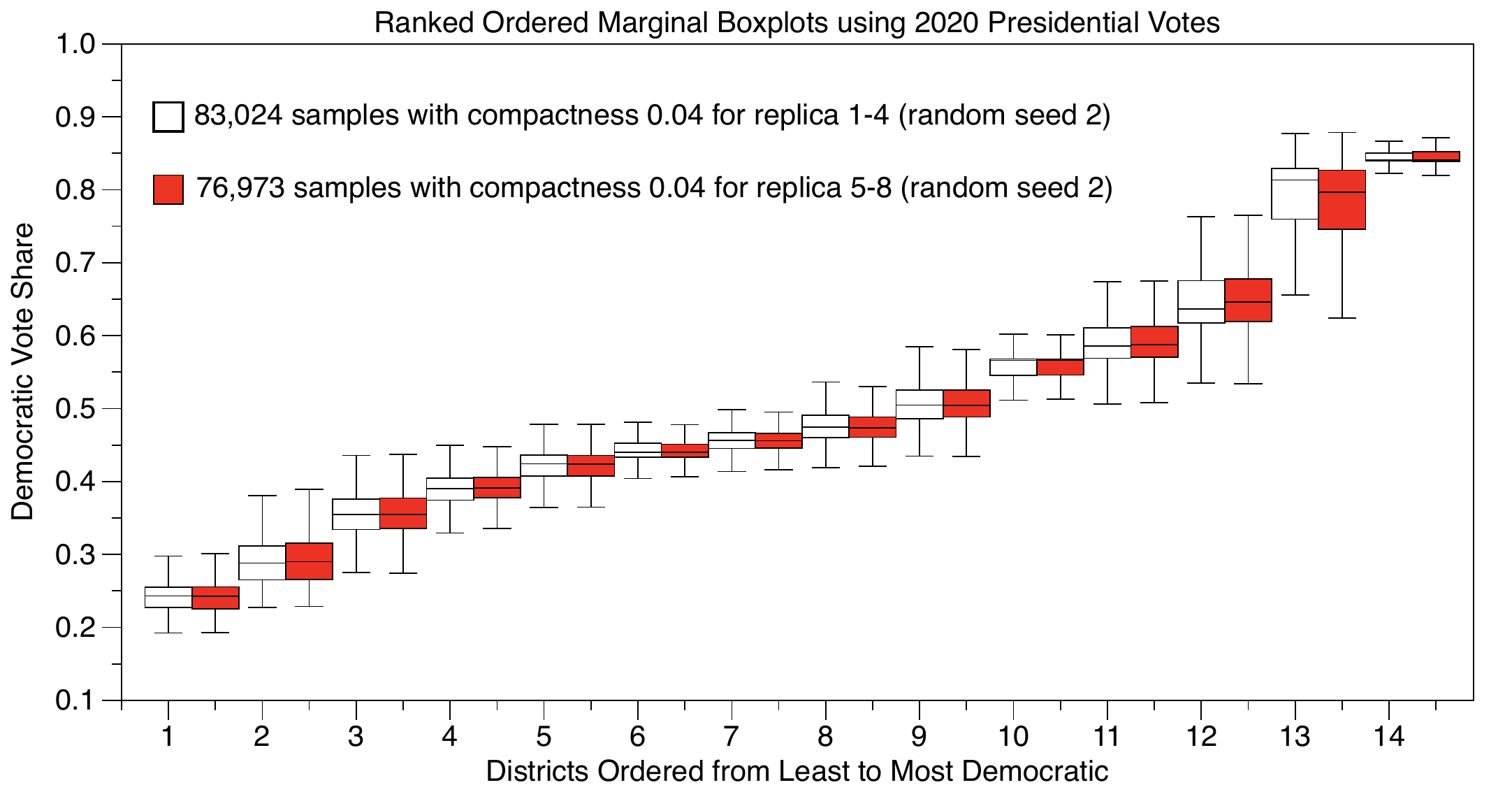}%
	\end{subfigure}
	    \caption{Comparison of the ranked 2020 presidential Democratic vote share in each district of each plan across two different sets of plans. The two sets of plans are taken from two sets of four different replicas from an eight-replica run of two million steps each, demonstrating that choice of replica was not significant.}
	\label{convergence2}%
\end{figure}

After confirming the i.i.d. samples in our heat bath, which ensures instantaneously mixing when swapping with the base measure, we then run the sampling method drawn from Equation (4) of the paper as summarized in the main body for up to four million steps under different random seeds. We use parallel tempering to get an appropriate  compactness weight with 11 replicas with compactness weight ranging from 0.00 to 0.1.\footnote{In tempering, we use the hyper-parameter $\gamma=0$ from the Metropolized Forest RECOM algorithm; see \cite{autrey2021metropolizednonmulti} for details.} To test for convergence, we compare the ensemble distribution from different random seeds, different running steps within the same random seed, and different replicas within the same random seeds and running steps, using the 2020 presidential election votes to see if the distributions mostly agree. See Figures~\ref{convergence1}, \ref{convergence3} and \ref{convergence2} for results. Note we show the convergence results of compactness weight $0.04$ since that is later found to be the closest compactness score to the 2021 map.

\section{Compactness}\label{compact_app}

For each district, the isoperimetric ratio is the square of its perimeter divided by its area. The total isoperimetric ratio of a districting plan is the sum of the isoperimetric ratios of each district. We use this total isoperimetric ratio as the measure of compactness in the parallel tempering process, with each interpolating distribution having a different weight~\cite{autrey2021metropolizednonmulti}.

\begin{figure}[H]
	\centering
	\includegraphics[width=0.7\linewidth]{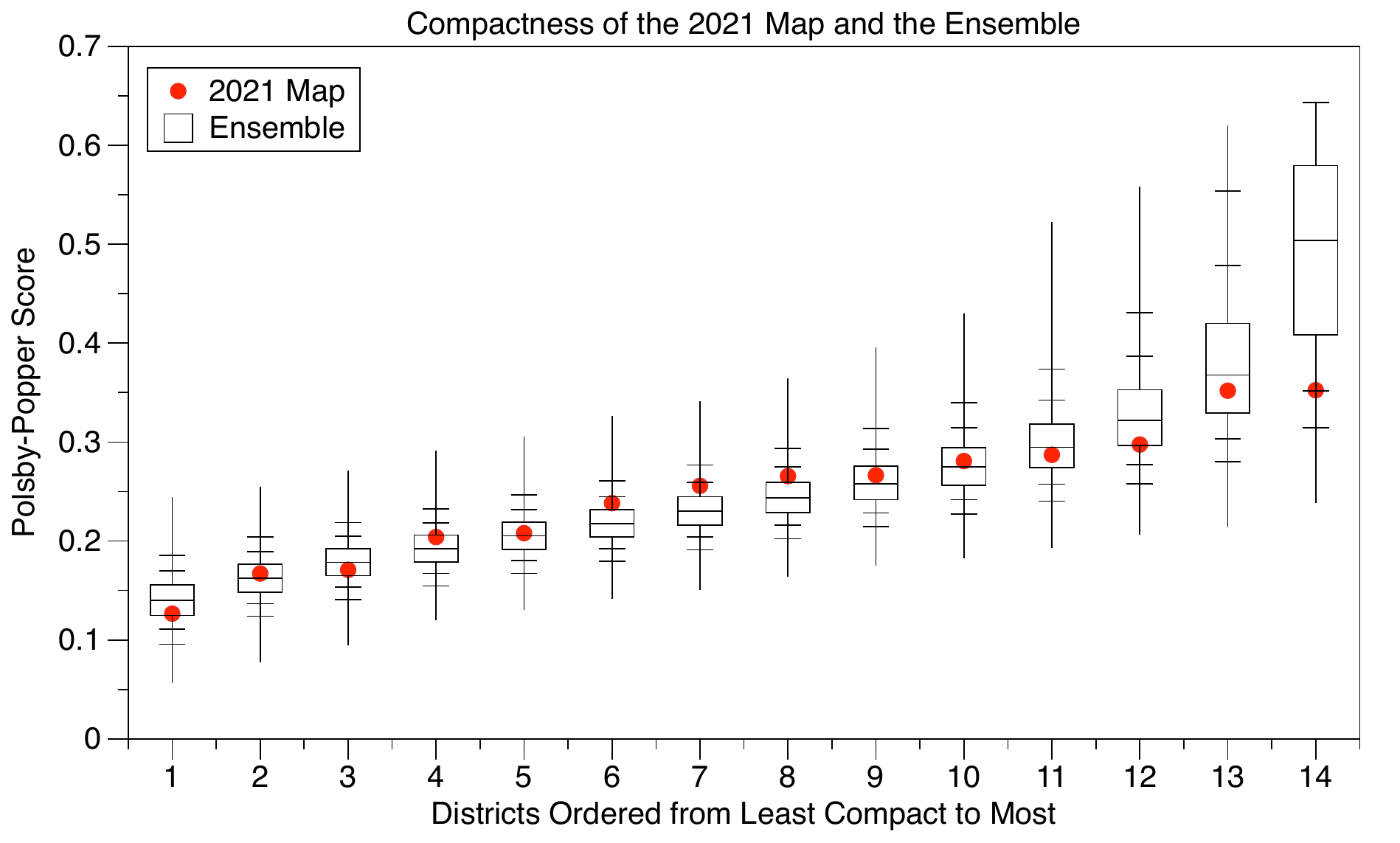}
	\caption{The red dots display the ordered Polsby-Popper score of the 14 districts in the enacted 2021 plan, compared to the ranges of Polsby-Popper scores observed for plans in the ensemble, ranked from least to most compact.
	}
	\label{compactness}
\end{figure}

To compare the ensemble with the enacted map, we use the Polsby-Popper score of each district, defined as $4\pi$ times the reciprocal of its isoperimetric ratio. Hence, a smaller isoperimetric ratio corresponds to larger Polsby-Popper scores. We observe that tempering up to compactness weight 0.04 produces an ensemble that converges (section~\ref{convergence_app}) and matches the 2021 map well, as measured by the ranked Polsby-Popper scores of the districts. See Figure~\ref{compactness} for this comparison.

\section{Additional uniform swing plots}\label{sec:app_swing}

\begin{figure}[h]%
	\centering
		\begin{subfigure}[t]{0.45\linewidth}
		\label{3a}%
		\includegraphics[width=\linewidth]{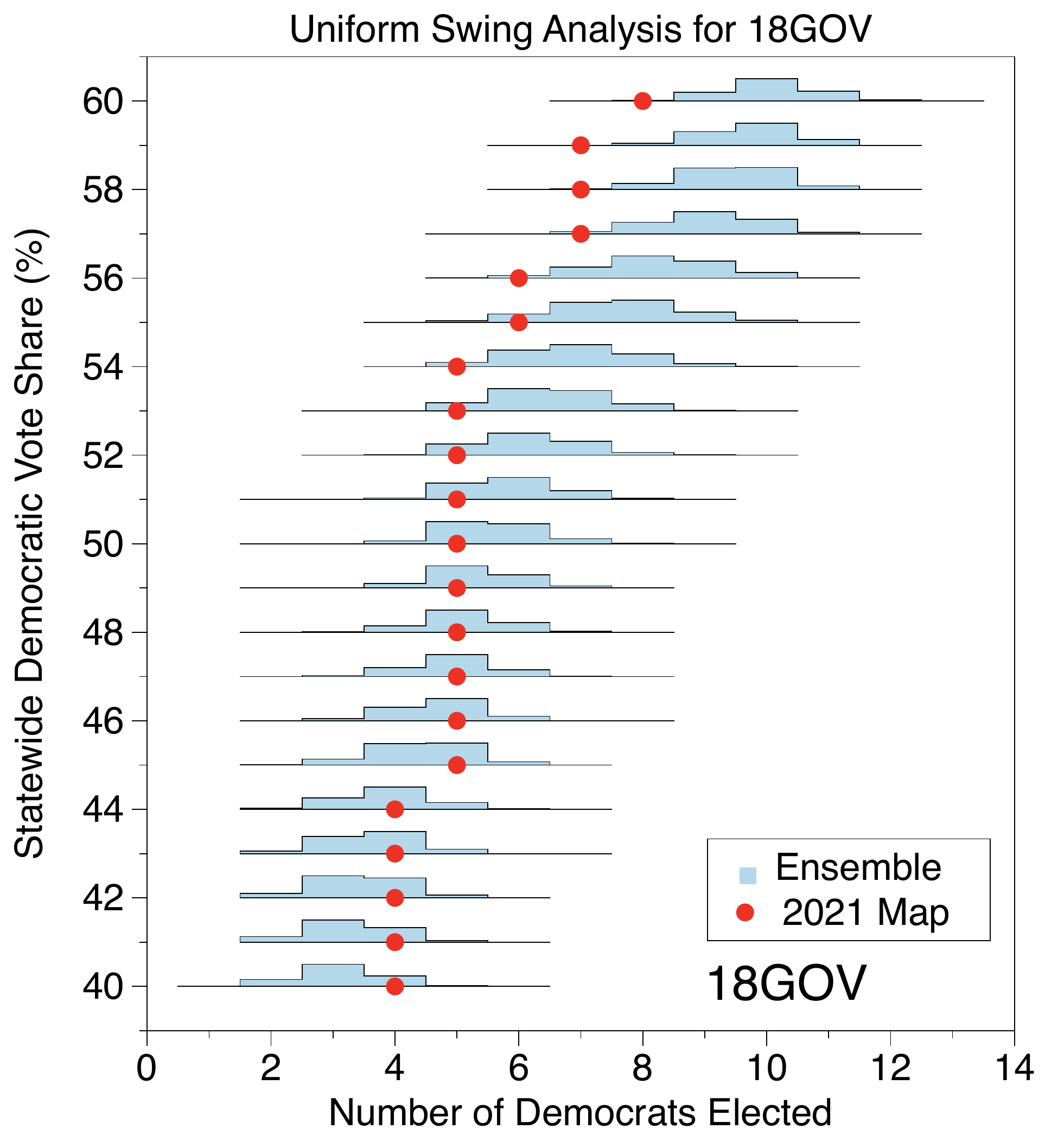}
    \end{subfigure}%
    	\hspace{18pt}%
    		\begin{subfigure}[t]{0.45\linewidth}
		\label{3b}%
		\includegraphics[width=\linewidth]{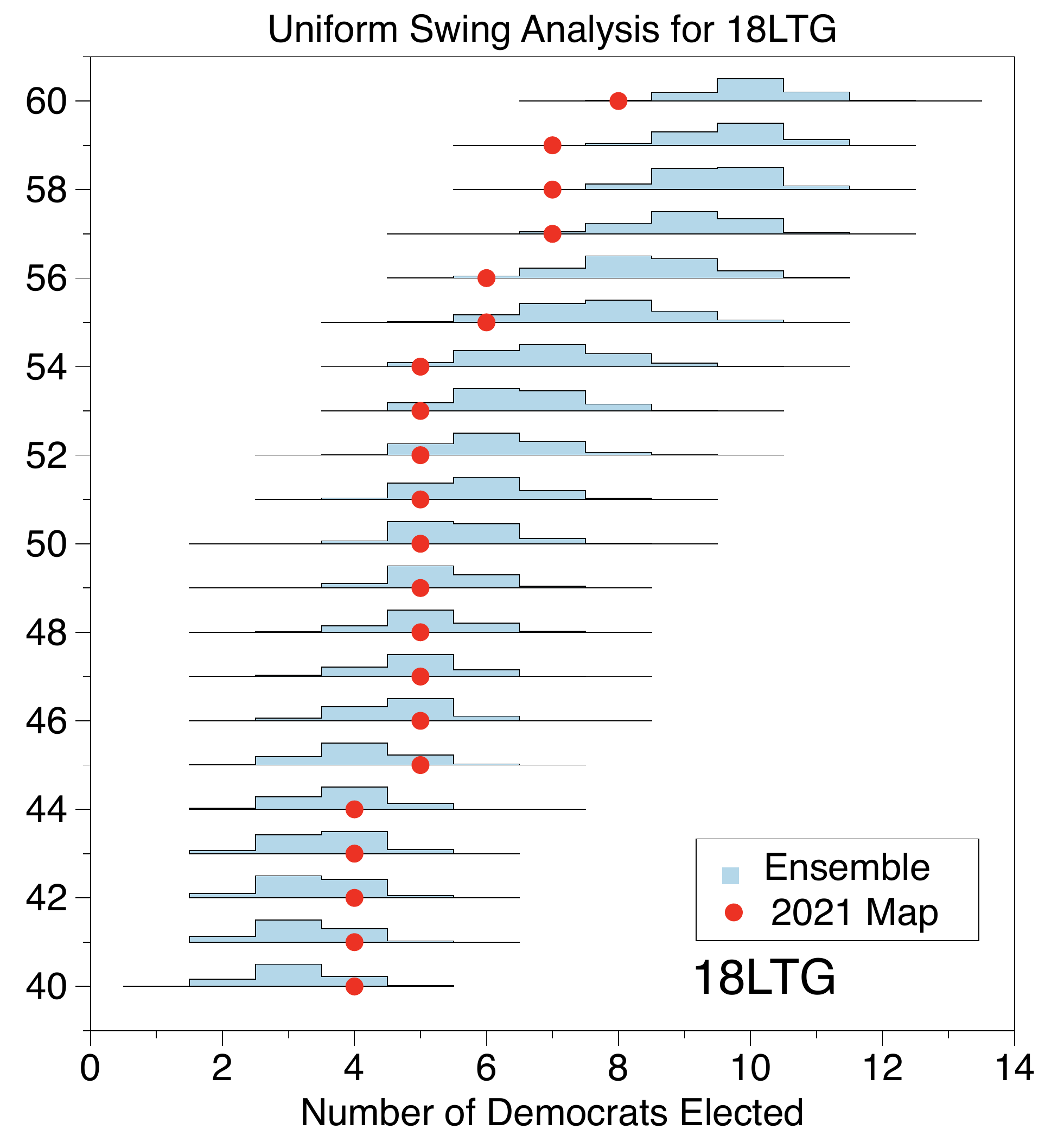}
    \end{subfigure}%
		\caption{The individual histograms give the frequency of the Democratic seat count in the ensemble for each of the shown statewide elections, with a uniform swing. The histograms are organized vertically based on the swung statewide partisan vote fraction. The red dots denote the Democratic seat count for the enacted plan for each of the swung vote profiles.}
	\label{swing18}%
\end{figure}

\begin{figure}[h]%
	\centering
	\begin{subfigure}[ht]{0.45\linewidth}
		\label{2a}%
		\includegraphics[width=\linewidth]{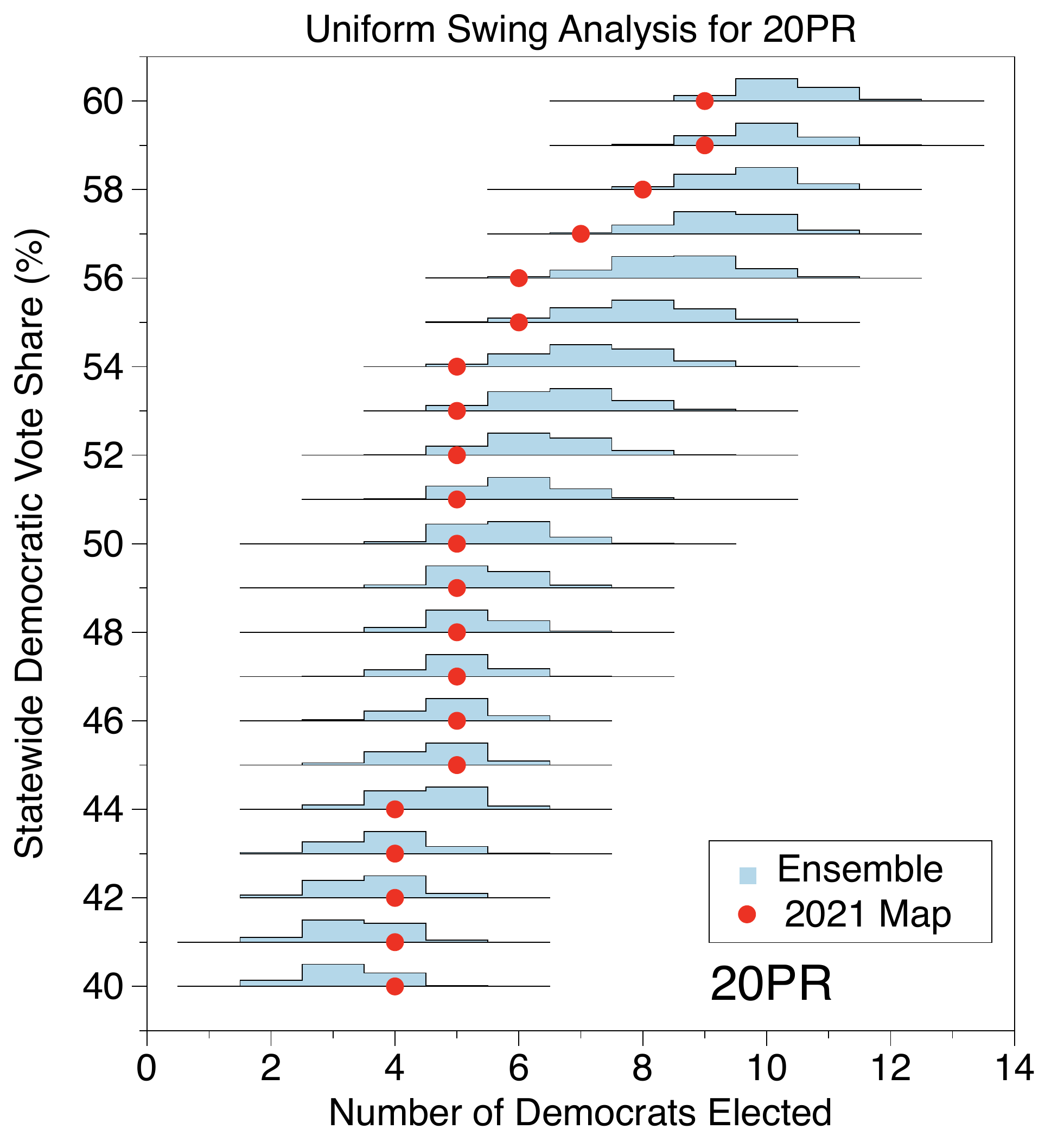}
    \end{subfigure}%
	\hspace{18pt}%
	\begin{subfigure}[ht]{0.45\linewidth}
		\label{2b}%
		\includegraphics[width=\linewidth]{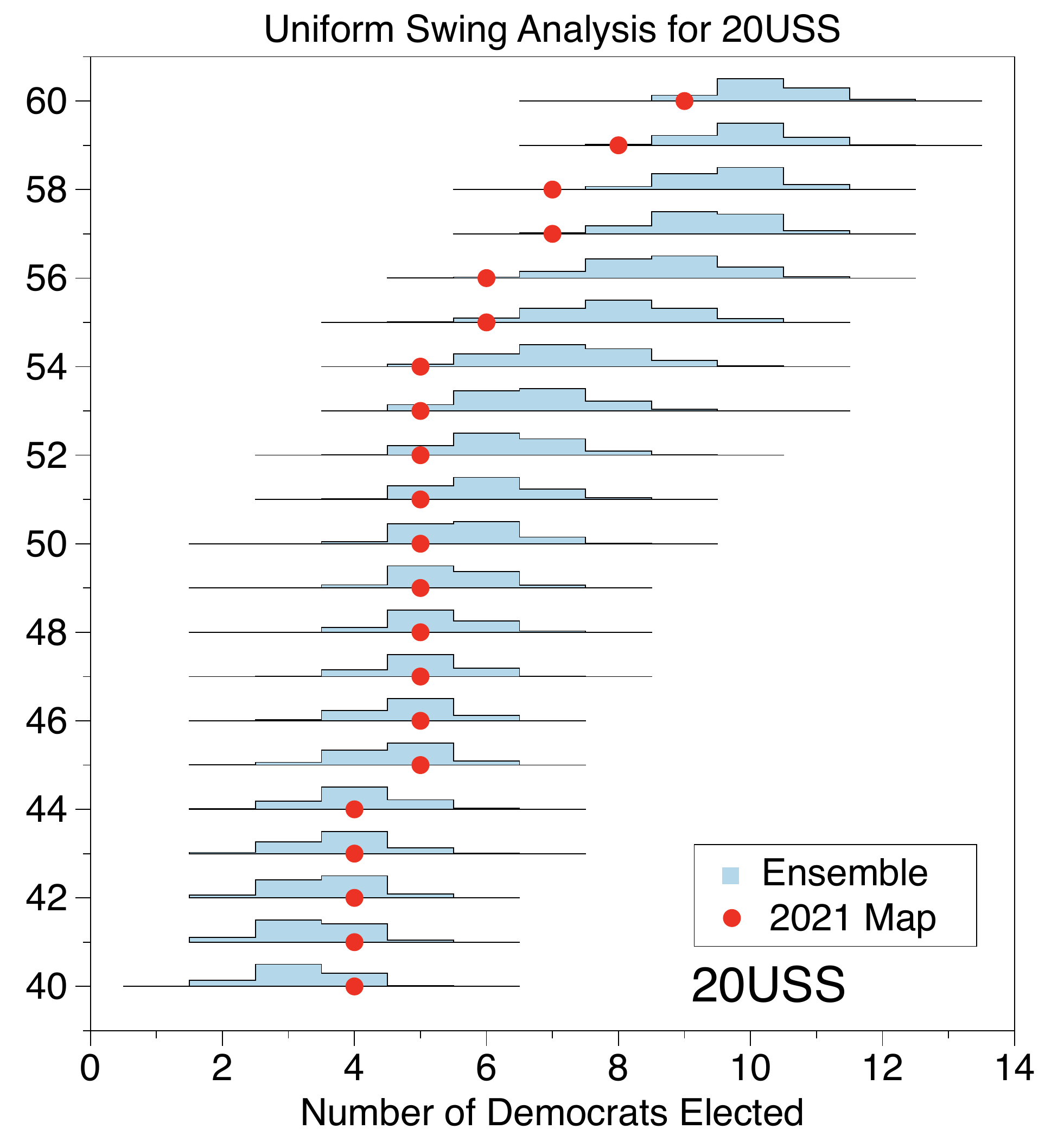}
    \end{subfigure}%
		\caption{The individual histograms give the frequency of the Democratic seat count in the ensemble for each of the shown statewide elections, with a uniform swing. The histograms are organized vertically based on the swung statewide partisan vote fraction. The red dots denote the Democratic seat count for the enacted plan for each of the swung vote profiles.}
	\label{swing20}%
\end{figure}

Figures~\ref{swing20} and~\ref{swing18} contain additional uniform swing analysis for 2018 and 2020 statewide elections. Again, as in Figure 3 of the paper, we observe that the enacted plan would result in nine Republican seats until the statewide Democratic vote share is swung to 55\%. When statewide Democratic votes are swung to range from 54\% to 60\%, the ensemble elects the same or fewer Democrats in  5.07\% for the 2018 gubernatorial election (18GOV); for the 2018 lieutenant gubernatorial election (18LTG), 4.55\%; for the 2020 presidential election (20PR), 8.91\%, and for the 2020 United States Senate election (20USS), 4.81\%.

\section{Rank-ordered marginal box plots}\label{sec:boxplots}

\begin{figure}[h]%
	\centering
	\begin{subfigure}[ht]{0.8\linewidth}
	    \label{6a}%
		\includegraphics[width=\linewidth]{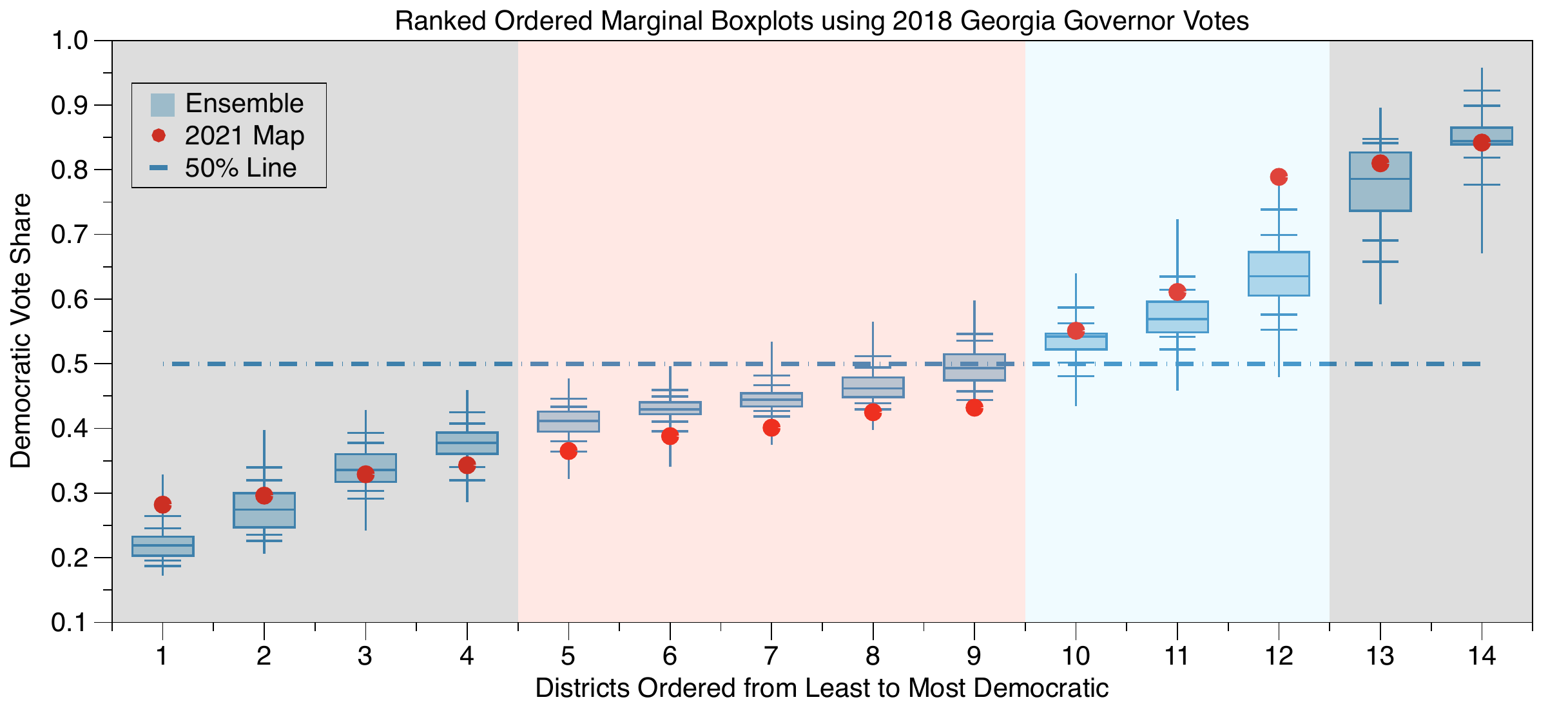}%
	\end{subfigure}
		\begin{subfigure}[ht]{0.8\linewidth}
	    \label{6b}%
		\includegraphics[width=\linewidth]{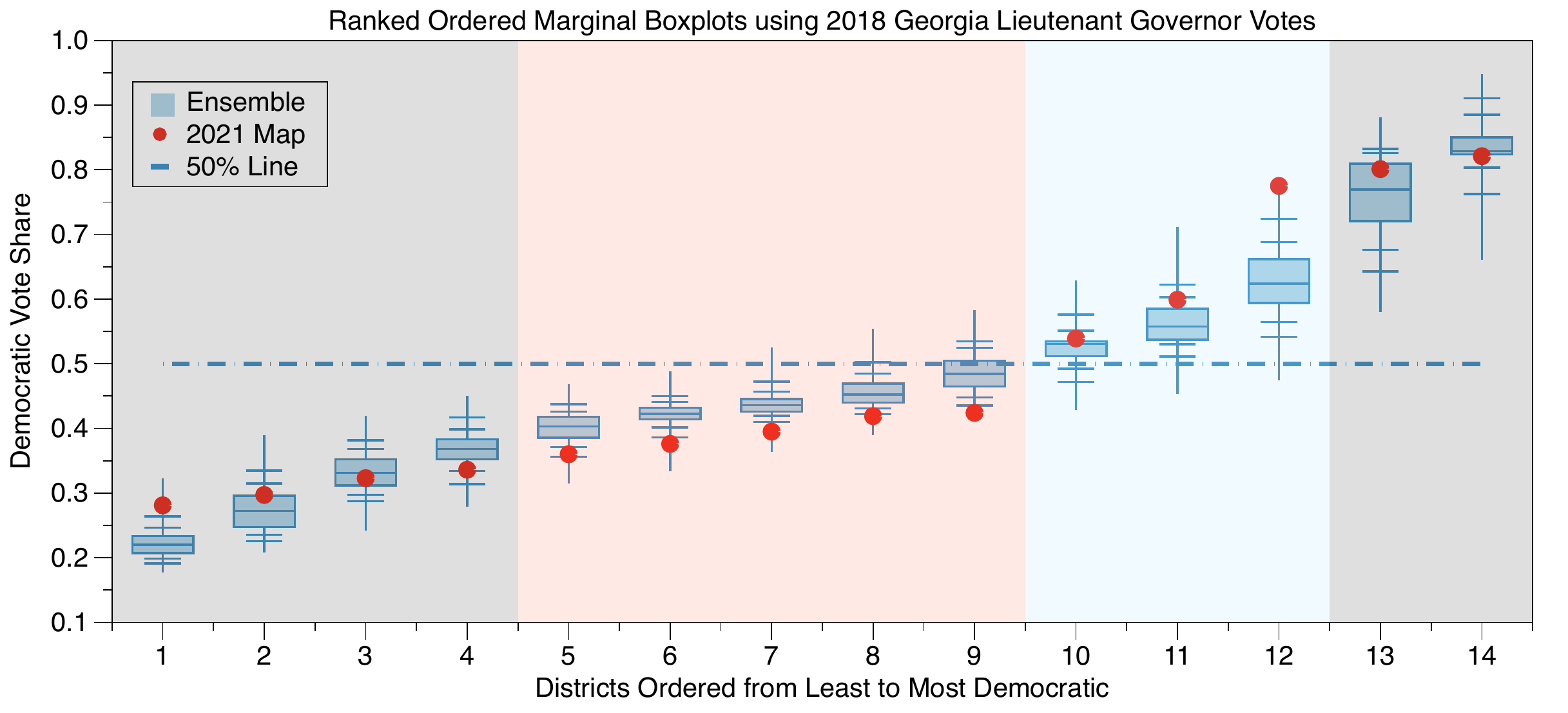}%
	\end{subfigure}
	\caption{Ranked Ordered Marginal box plots using Georgia Governor and Lieutenant Governor 2018 votes.
	}
	\label{box18}%
\end{figure}

Figure~\ref{box18} contains additional rank-ordered marginals for 2018 statewide elections which are typical of the larger range of 2018 elections we consider. As in Figures 4 and 5 of the paper, we observe greatly increased polarization in the enacted map, with the $5^{\rm{th}}$-$9^{\rm{th}}$ most Republican districts having far more Republicans than is typical in the ensemble and the $10^{\rm{th}}$-$12^{\rm{th}}$ most Republican districts having far more Democrats than is typical in the ensemble. In Table~\ref{tab:polarization}, we display the number of plans in the ensemble with the same or fewer Democratic voters in the $5^{\rm{th}}$-$9^{\rm{th}}$ most Republican districts and the number of plans in the ensemble with the same or more Democratic voters in the $10^{\rm{th}}$-$12^{\rm{th}}$ most Republican districts. 

\begin{table}\footnotesize
\begin{tabularx}{\columnwidth} { 
  | >{\centering\arraybackslash}X
  | >{\centering\arraybackslash}X 
  | >{\centering\arraybackslash}X| }
 \hline
 Elections  & $\%$Ensemble plans with same or \textbf{fewer} Dem. votes than 2021 enacted map in the $5^{\rm{th}}$-$9^{\rm{th}}$ most Rep. districts  & $\%$Ensemble plans with same or \textbf{more} Dem. votes than 2021 enacted map in the $10^{\rm{th}}$-$12^{\rm{th}}$ most Rep. districts \\
 \hline
20PR    & $0.007\%$ (11/159,997)    &  $0.200\%$ (320/159,997) \\
\hline
20USS   & $0.000\%$ (0/159,997)   &  $0.158\%$ (252/159,997) \\
\hline
20PSC1    & $0.000\%$ (0/159,997)    &  $0.078\%$ (124/159,997)   \\
\hline
20PSC4   & $0.000\%$ (0/159,997)    &  $0.085\%$ (136/159,997) \\
\hline
18GOV   & $0.000\%$ (0/159,997)    &  $0.154\%$ (246/159,997) \\
\hline
18LTG    & $0.000\%$  (0/159,997)    &  $0.144\%$ (231/159,997)  \\
\hline
18ATG  & $0.000\%$ (0/159,997)   &  $0.088\%$ (140/159,997)  \\
\hline
18LAB    & $0.000\%$ (0/159,997)     &  $0.086\%$ (138/159,997)  \\
\hline
18AGR    & $0.000\%$ (0/159,997)    &  $0.082\%$ (131/159,997) \\
\hline
18PSC3   & $0.000\%$ (0/159,997)   &  $0.109\%$ (174/159,997) \\
\hline
18PSC5    & $0.000\%$ (0/159,997)   &  $0.082\%$ (131/159,997) \\
\hline
18SOS   & $0.001\%$ (1/159,997)     &  $0.091\%$ (146/159,997) \\
\hline
18SOSro   & $0.017\%$ (27/159,997)    &  $0.129\%$ (207/159,997) \\
\hline
18INS    & $0.000\%$ (0/159,997)   &  $0.084\%$ (134/159,997) \\
\hline
18SPI    & $0.000\%$ (0/159,997)   &  $0.069\%$ (107/159,997) \\
\hline
16USS   & $0.000\%$ (0/159,997)    &  $0.016\%$ (25/159,997) \\
\hline
16PR    & $0.000\%$ (0/159,997)    &  $0.049\%$ (78/159,997) \\
\hline
\end{tabularx}
    \caption{\small{Starting from the left, the first column gives the statewide election whose abbreviation is given in Figure 2 of the paper.The total number of plans in the ensemble considered is 159,997. The second column gives the percentage of the number of the plans out of those 159,997 plans which have the same or \textbf{fewer} Democratic votes than the enacted 2021 Map for the $5^{\rm{th}}$-$9^{\rm{th}}$-most Republican districts. 
    These extremely low percentages show that the $5^{\rm{th}}$-$9^{\rm{th}}$-most Republican districts in the enacted plan are significantly more Republican than they typically are in the ensemble.
    The third column gives the percentage of the number of the plans out of those 159,997 plans which have the same or \textbf{more} Democratic votes than the enacted 2021 Map for the $10^{\rm{th}}$-$12^{\rm{th}}$ most Republican districts.
    These extremely low percentages show that the $10^{\rm{th}}$-$12^{\rm{th}}$-most Republican districts in the enacted plan are significantly more Democratic than they typically are in the ensemble.
    Together, these demonstrate the strong polarization of the enacted plan.}}
    \label{tab:polarization}
\end{table}

\section{The 2021 plan and the 2011 plan}
\begin{figure}
    \centering
    \includegraphics[width=0.38\linewidth]{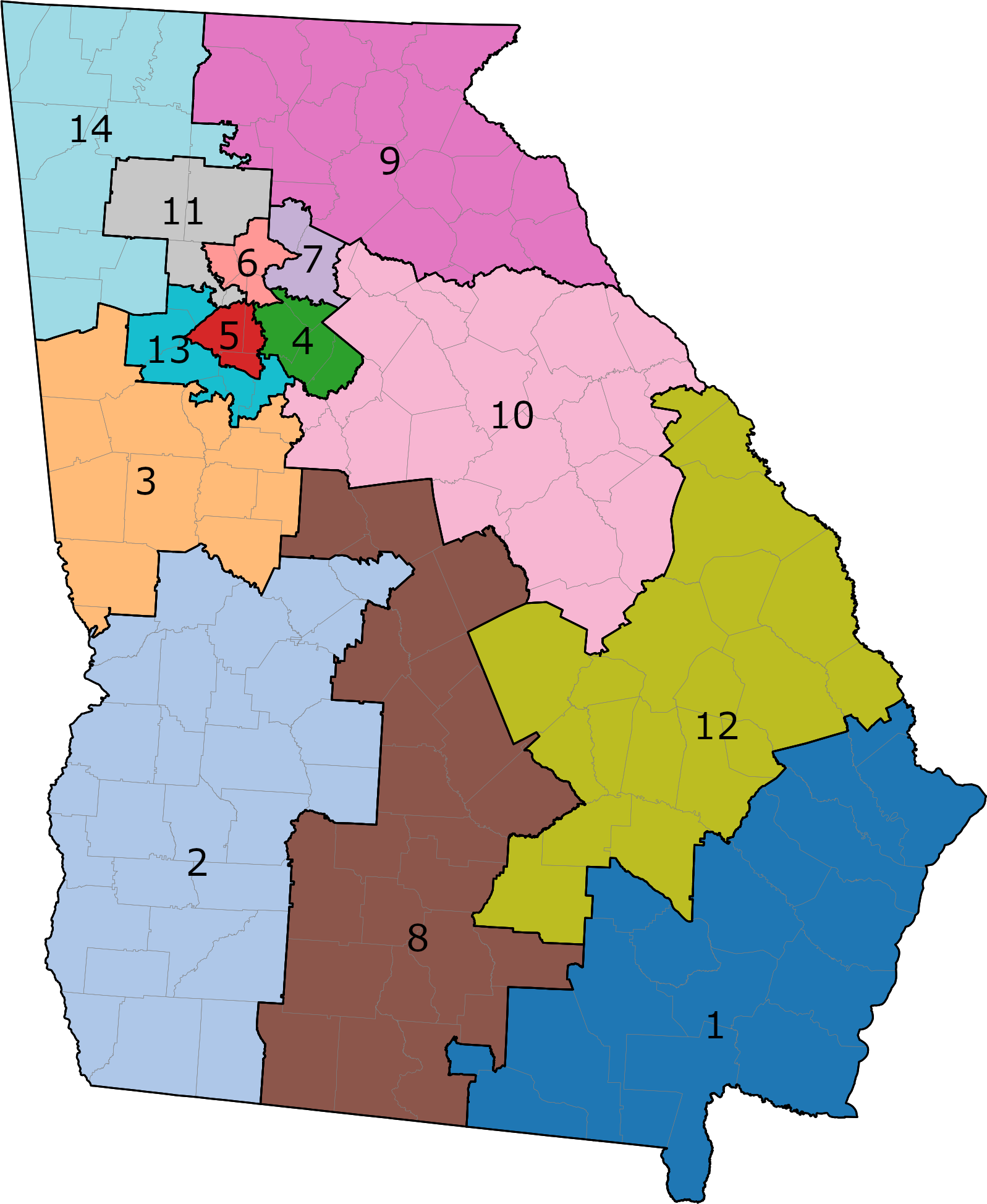}
    \includegraphics[width=0.38\linewidth]{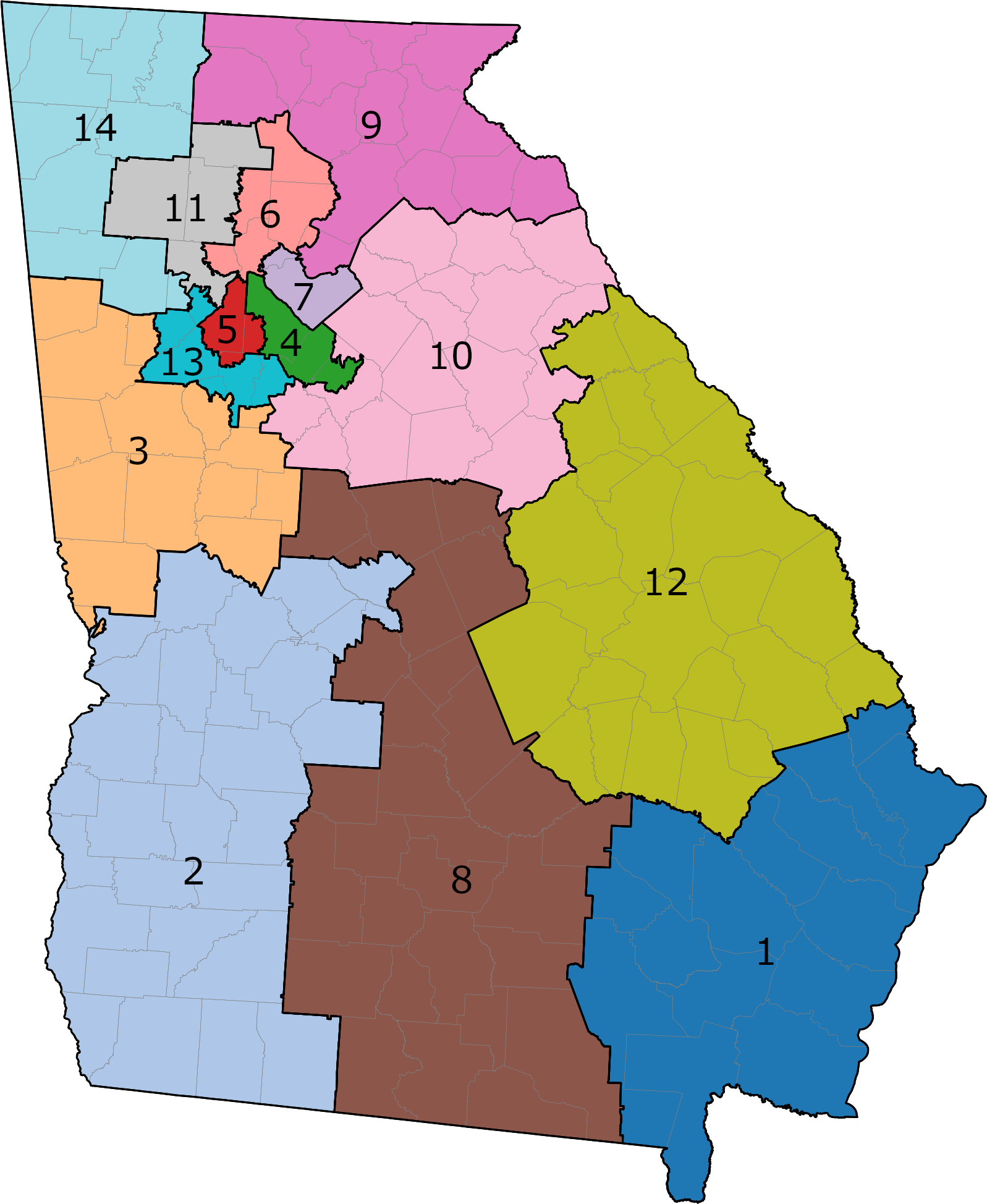}
    \caption{The 2011 (left) and 2021 (right) congressional plans, with each district labelled by its name.}
    \label{fig:2011vs2021enacted}
\end{figure}

Figure~\ref{fig:2011vs2021enacted} shows the 2011 plan and 2021 plan. Comparing both plans, we can find that there is close alignment between the overarching structures of the 2011 and 2021 district plans. For example, the 1st District in 2011 “looks like” the 1st District in 2021. 

\section{The Voting Rights Act}\label{asec:vra}
The enacted 2021 plan in Georgia contains four districts that either are or nearly are majority black districts in terms of the voting age population. These four districts will allow for black voters to elect candidates of their choice, and these four representatives will proportionally represent the black population of Georgia. However, over the past two decades, both older and newer established research on the Voting Rights Act (VRA) reveals that minority-electing districts do not need to be majority-minority \cite{grofman2000drawing,becker2021computational}.  In fact, a minority voting block may comprise significantly fewer than half of the eligible voters and still be capable of electing a representative of their choice.
This observation is supported by the recent 2017 US Supreme Court ruling in \emph{Cooper v. Harris} \cite{2016cooperharris}.  In this case, North Carolina congressional districts were ruled to be racial gerrymanders because they over-packed black voters into two majority-black districts, thereby diluting these voters' influence in surrounding districts.

In this work, we generated our ensemble in a way that is agnostic to generating minority-electing districts. From this ensemble, we concluded that the lack of responsiveness in the enacted plan was atypical in redistricting.  However, it is possible that the lack of responsiveness instead derives from the lack of a VRA consideration rather than political considerations. To test this, we filter our ensemble to only consider districts that have four VRA districts with the power to elect minority candidates.  We use the methods of Grofman et al. \cite{grofman2000drawing} to filter our ensemble and investigate the results of enforcing four districts in which black voters have a ``good'' chance of electing a Representative of their choice.

In this section, we show that over half of the plans in our ensemble contain  four districts that could elect a black representative. Sub-sampling these districts reveals no qualitative changes in our results.  We find evidence that enforcing VRA districts, as defined above, will not qualitatively change our conclusions, meaning that the pursuit of enforcing the VRA is unlikely to cause of the observed lack of responsiveness.

In this analysis, we search for plans in which black voters have a reasonable chance at winning both the primary and general elections with a candidate of their choice.  A challenge for us in this study is that we do not have precinct level primary data in Georgia.  We instead examine the 17 elections mentioned in Table~\ref{tab:polarization}.  Of these elections, three of them had a black candidate running for Democratic office (18GOV, 18INS, and 20USS) and we pay special attention to these three elections.  

We base our analysis off of a simple model from~\cite{grofman2000drawing}. The model assumes a ``worst case" scenario for black voters in primary elections in which (i) black voters vote as a block in support of a Democratic candidate and (ii) non-black Democratic voters will exclusively vote against the black-preferred primary candidate.  Formally, we require a VRA district has the following property
\begin{align}
    2B_{V} &> D_{V},\\
    B_{V} &= B_{VAP} \frac{D_{V} + R_{V}}{T_{VAP}},
\end{align}
where $B_{V}$ represents the black voter turn out, $D_V$ and $R_V$ represent the number of Democratic and Republican votes, respectively, and $B_{VAP}$ and $T_{VAP}$ represents the black and total voting age populations, respectively.  

Victory in the general election is model with a parameter $c$ which reflects the non-black Democratic voters who vote with the black preferred candidate in a general election.  Formally we require that ,
\begin{align}
    B_{V} + c(D_{V}-B_V) &> R_{V} + (1-c)(D_{V}-B_V).
\end{align}
$c$ is, in general, an unknown parameter.  To estimate it we compare each of the 14 elections we have in which there was no black candidate to each of the 3 elections we have in which their was a black candidate. We then assume that the Democratic vote fraction in the election without a black candidate would have matched the Democratic vote fraction in the election with a black candidate, before accounting for racially polarized voting.  Formally, this means that 
\begin{align}
D_{V}^{(b)} &= B_V^{(b)} + c(D_{V}^{(nb)'} - B_V^{(b)}),\\
\frac{D_{V}^{(nb)'}}{D_V^{(b)} + D_V^{(b)}} &= \frac{D_{V}^{(nb)}}{D_V^{(nb)} + D_V^{(nb)}} 
\end{align}
where the superscripts $(b)$ and $(nb)$ refer to elections with and without a black candidate respectively.  $D_{V}^{(nb)'}$ represents the votes that \emph{would} have gone to the Democratic candidate if they were not black.  The second equation assumes that the fraction of the democrative vote under the observed election with no black candidates matches the hypothetical election that would lead to $D_{V}^{(nb)'}$ Democratic votes.  We can then determine $c$ from the model by comparing different combinations of elections.

We find that $c=1.07\pm0.14$, which reveals that voters in Georgia typical vote based on party rather than race in general elections.  We also test for local effects on $c$ by restricting our analysis to the four enacted congressional districts that largest black voting age populations and also the complement of this set.  In both alternative cases, we again found that $c$ is very close to one, supporting the idea that which reveals that voters in Georgia typically vote based on party independent of where they are in the state.

Choosing $c=1$ reduces to the condition that the Democratic vote is greater than the Republican vote.  We remark that this is a fairly conservative model, as demonstrated in \cite{becker2021computational}.

\begin{figure}[t]%
\centering
	\includegraphics[width=0.95\linewidth]{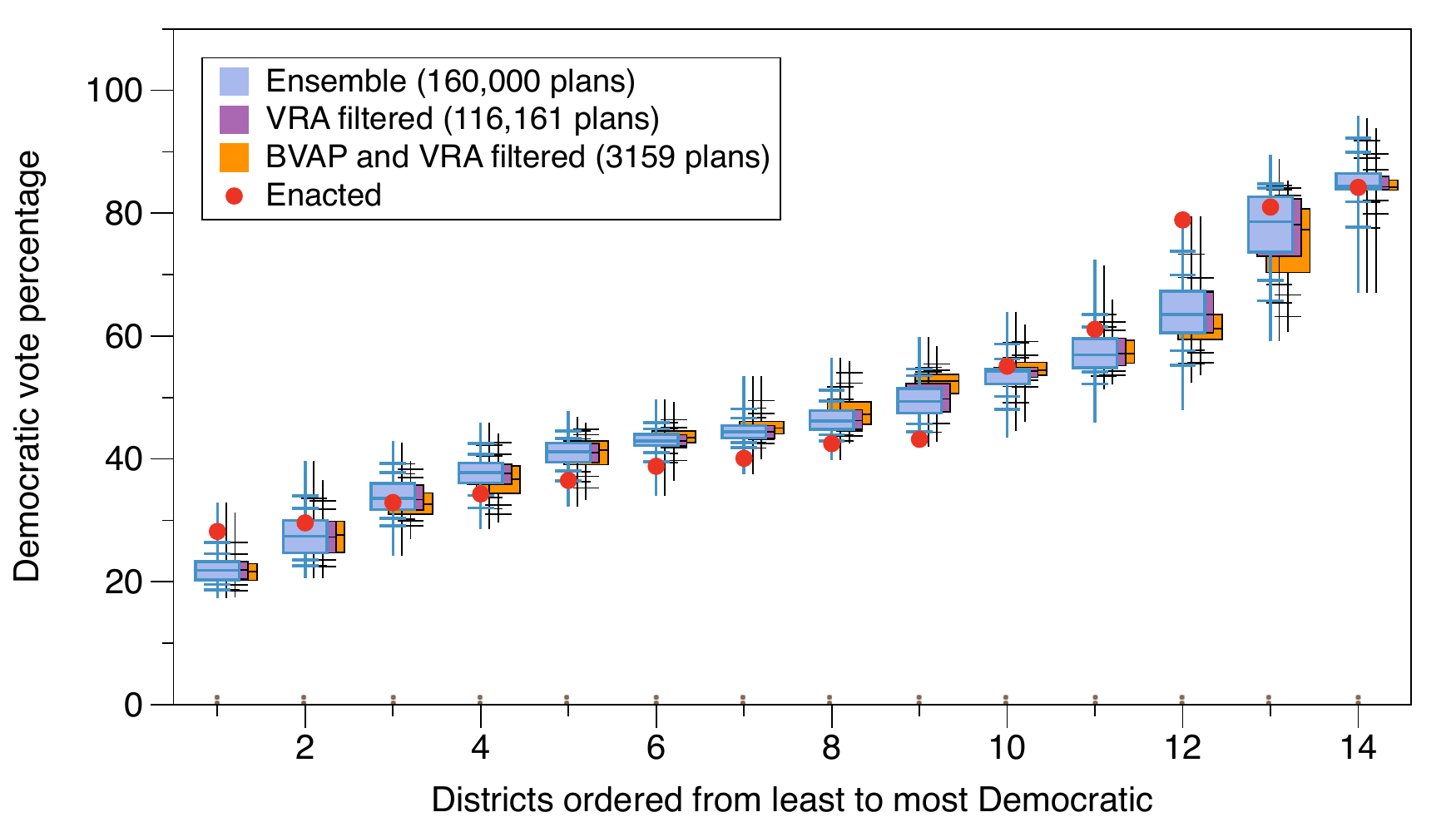}	
	\caption{We demonstrate the impact of filtering on VRA districts and on VRA districts that have a high black voting age population. In both cases we do not see significant shifts that could explain the behavior of the enacted plan.}
\label{fig:VRAbox}
\end{figure}

Having fixed criteria for VRA districts, we examine the number of plans in which at least fourdistricts satisfy the primary and general election constraints in at least 14 of the 17 elections. We also ensure that the four constraints are met in the same four districts under all three of the elections with black candidates. We remark that in the recent, more nuanced analysis of \cite{becker2021computational}, the authors seek to create VRA districts which have a 60\% chance of electing the minority candidate of choice; in the current work our 14 of 17 elections corresponds to over an 80\% chance.  We find that only 116,161 of our 160,000 plans (roughly 80\% of our plans) satisfy these constraints.
We validate that the removal of 20\% of the remaining plans do not significantly shift our ensemble in Figure~\ref{fig:VRAbox}.

In both the literature \cite{grofman2000drawing, becker2021computational} and in the US Supreme Court \cite{2016cooperharris}, establishing majority-minority districts is not necessary for establishing VRA districts.  In fact, in \cite{2016cooperharris}, the courts ruled that majority-minority districts may serve to dilute the voice of minorities in surrounding districts by over representing them in the VRA district. 

Nevertheless, Georgia's congressional plan contains four (nearly) majority-minority districts.  To test the hypothesis that the generation of majority-minority districts is responsible for the lack of responsiveness, we add an additional filter to the 116,161 VRA-filtered plans by only examining plans in which the at least four of the candidate VRA districts also have at least 45\% black voting age population.  We find that only 3159 of the 116,161 plans satisfy this additional constraint.  

Although these 3159 plans do not provide a representative sample of plans drawn to generate majority-minority districts, we can still treat them as random samples drawn from such a distribution and investigate potential bias that may occur due to this further restriction.  We examine the box plots of these districts also in Figure~\ref{fig:VRAbox} and find no significant shift that would explain the extreme behavior of the enacted plan; in fact, we see a reduction in the Democratic vote fractions in the most Democratic districts, which would exacerbate the relative extremity of the enacted plan.

\begin{figure}[h]%
	\centering
	\begin{subfigure}[h]{.49\linewidth}%
		\includegraphics[width=\linewidth]{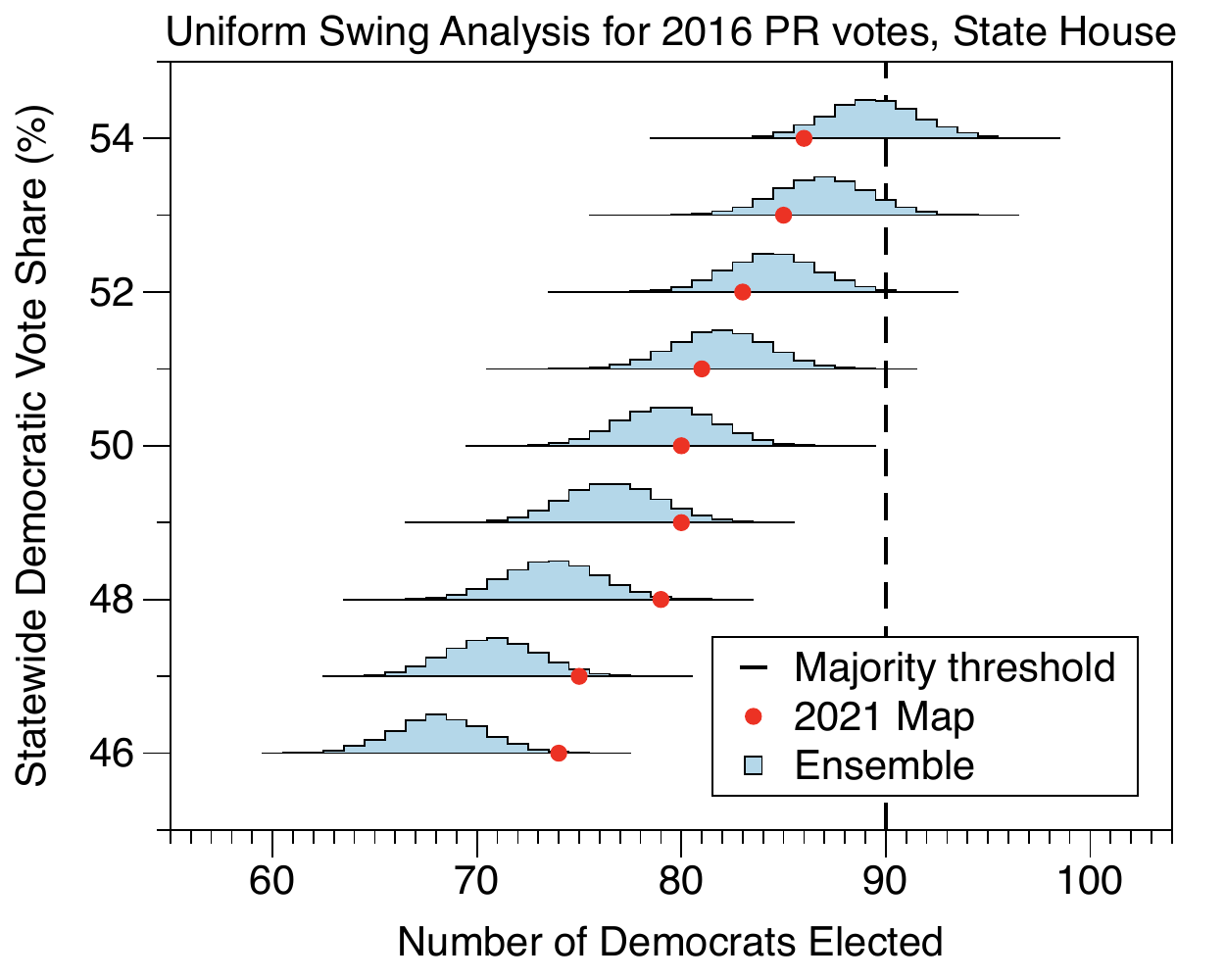}%
	\end{subfigure}
		\centering
	\begin{subfigure}[h]{.49\linewidth}%
		\includegraphics[width=\linewidth]{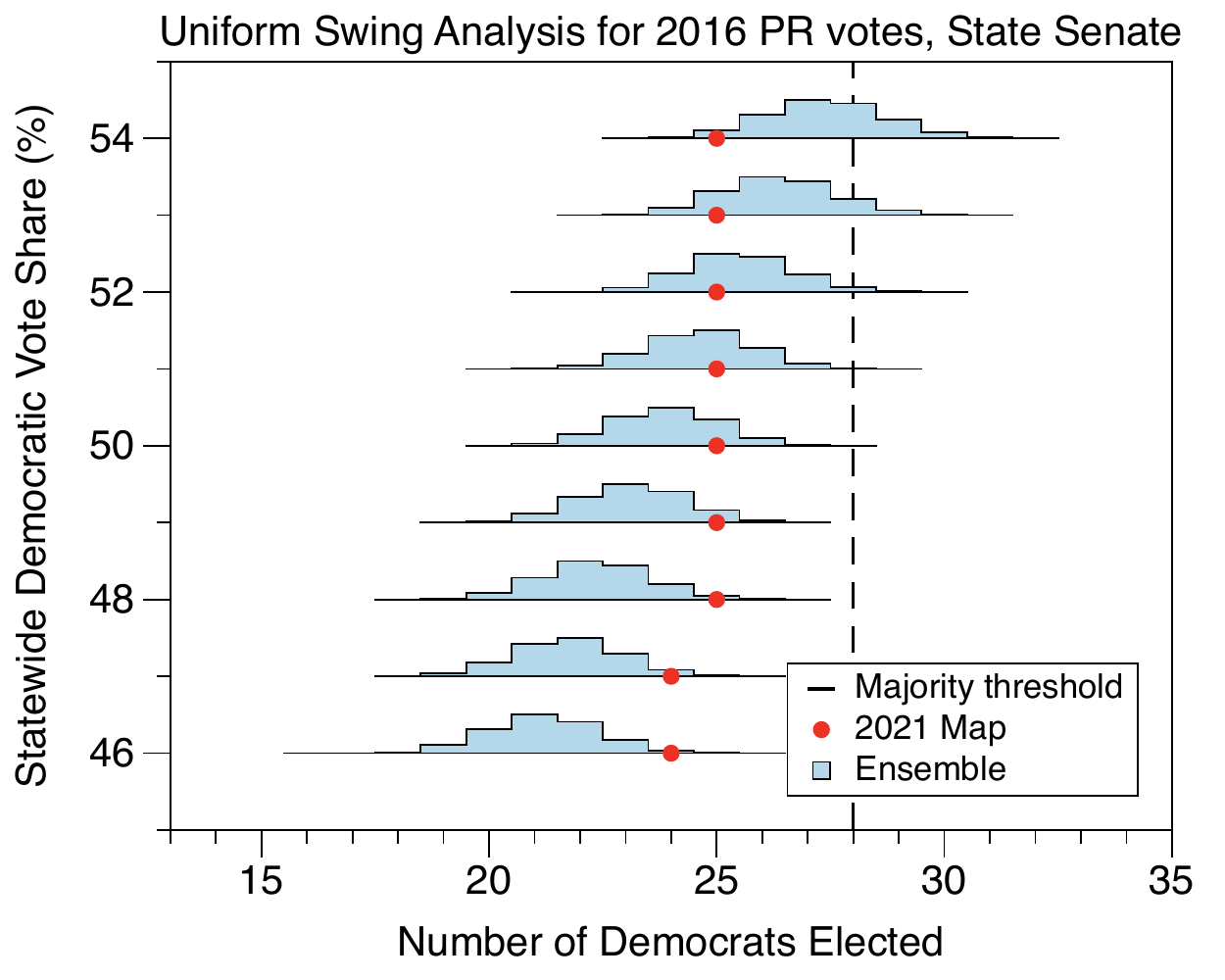}%
	\end{subfigure}
		\begin{subfigure}[h]{.49\linewidth}%
		\includegraphics[width=\linewidth]{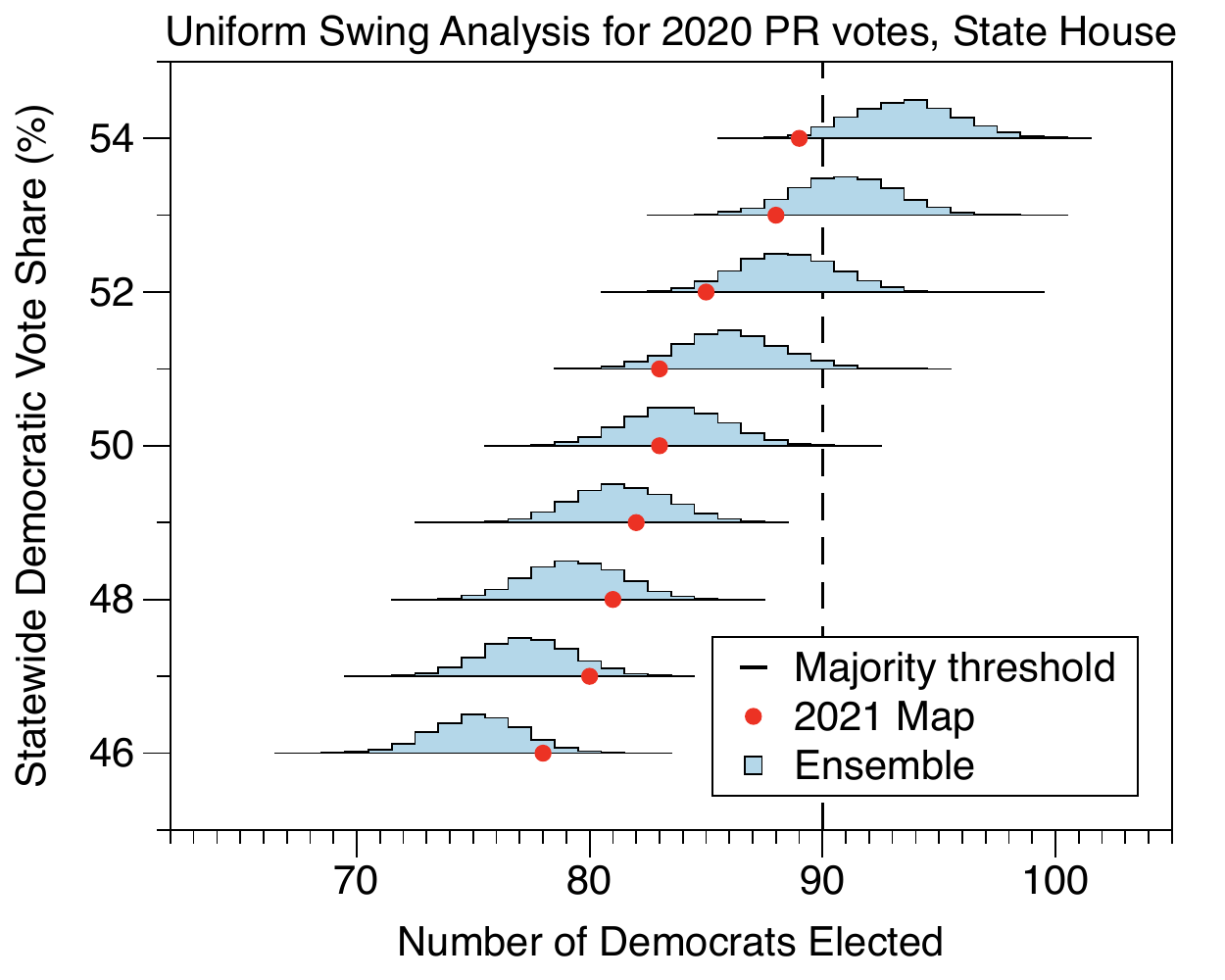}%
\end{subfigure}
		\begin{subfigure}[h]{.49\linewidth}%
		\includegraphics[width=\linewidth]{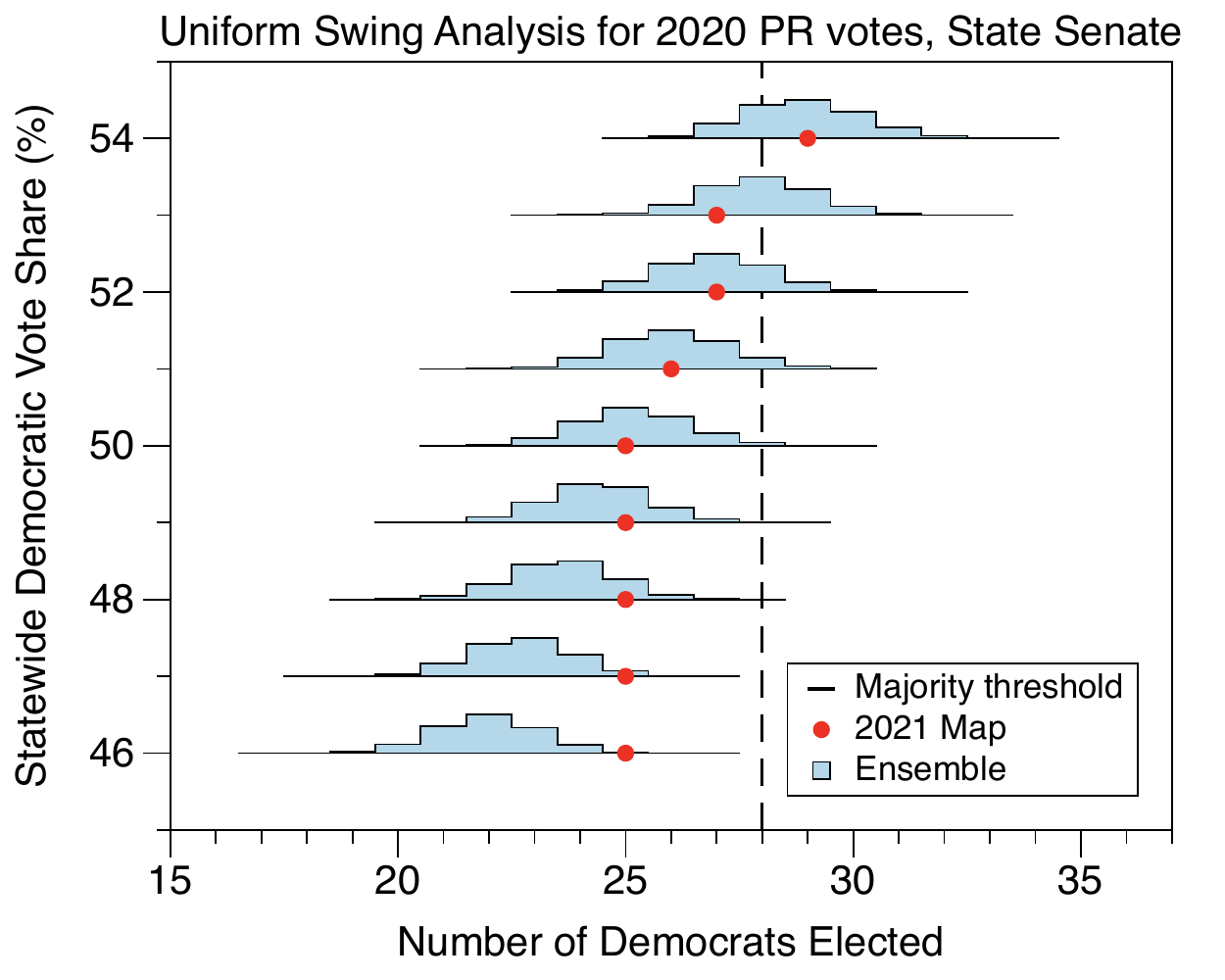}%
	\end{subfigure}
			\begin{subfigure}[h]{.49\linewidth}%
		\includegraphics[width=\linewidth]{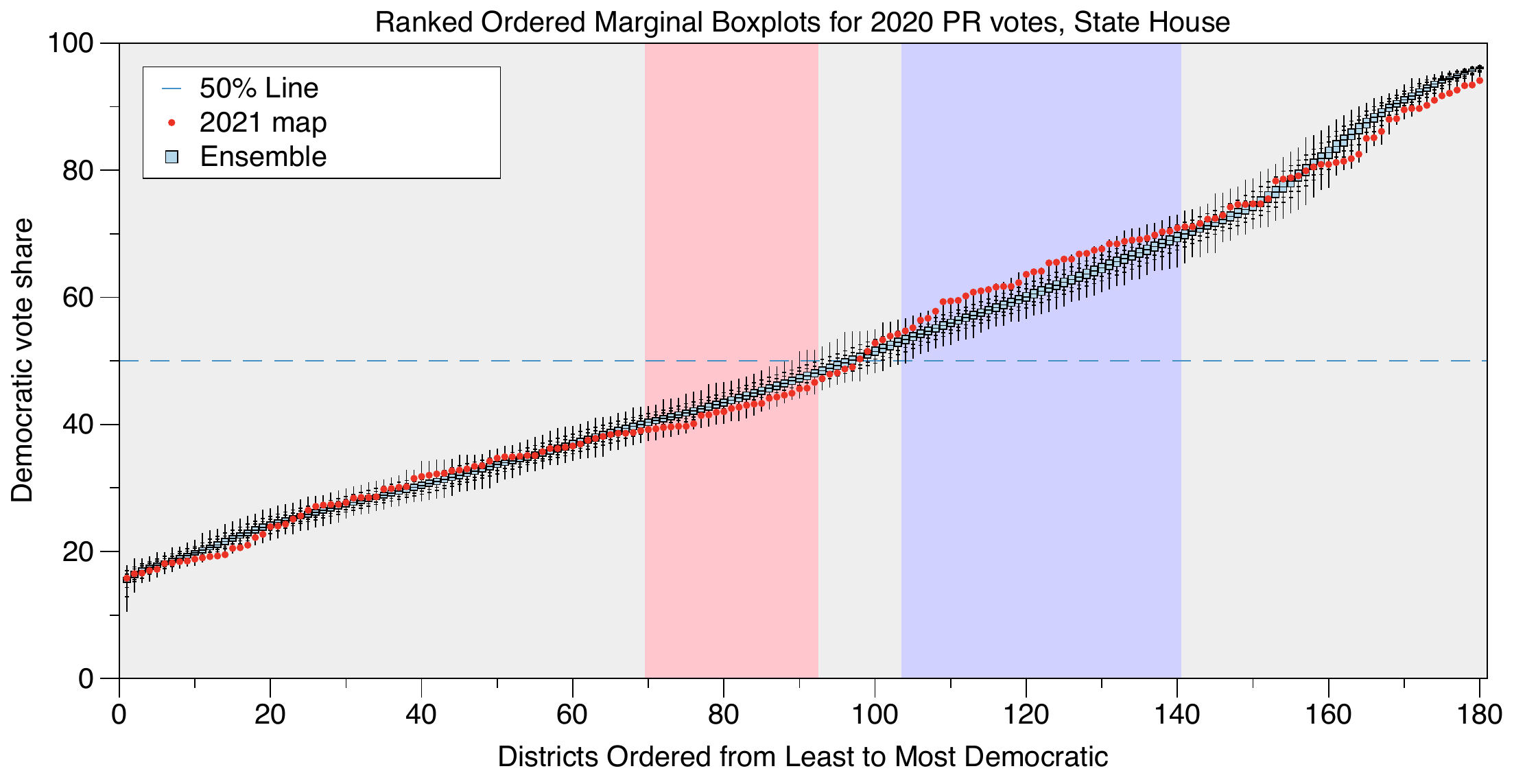}%
	\end{subfigure}
				\begin{subfigure}[h]{.49\linewidth}%
		\includegraphics[width=\linewidth]{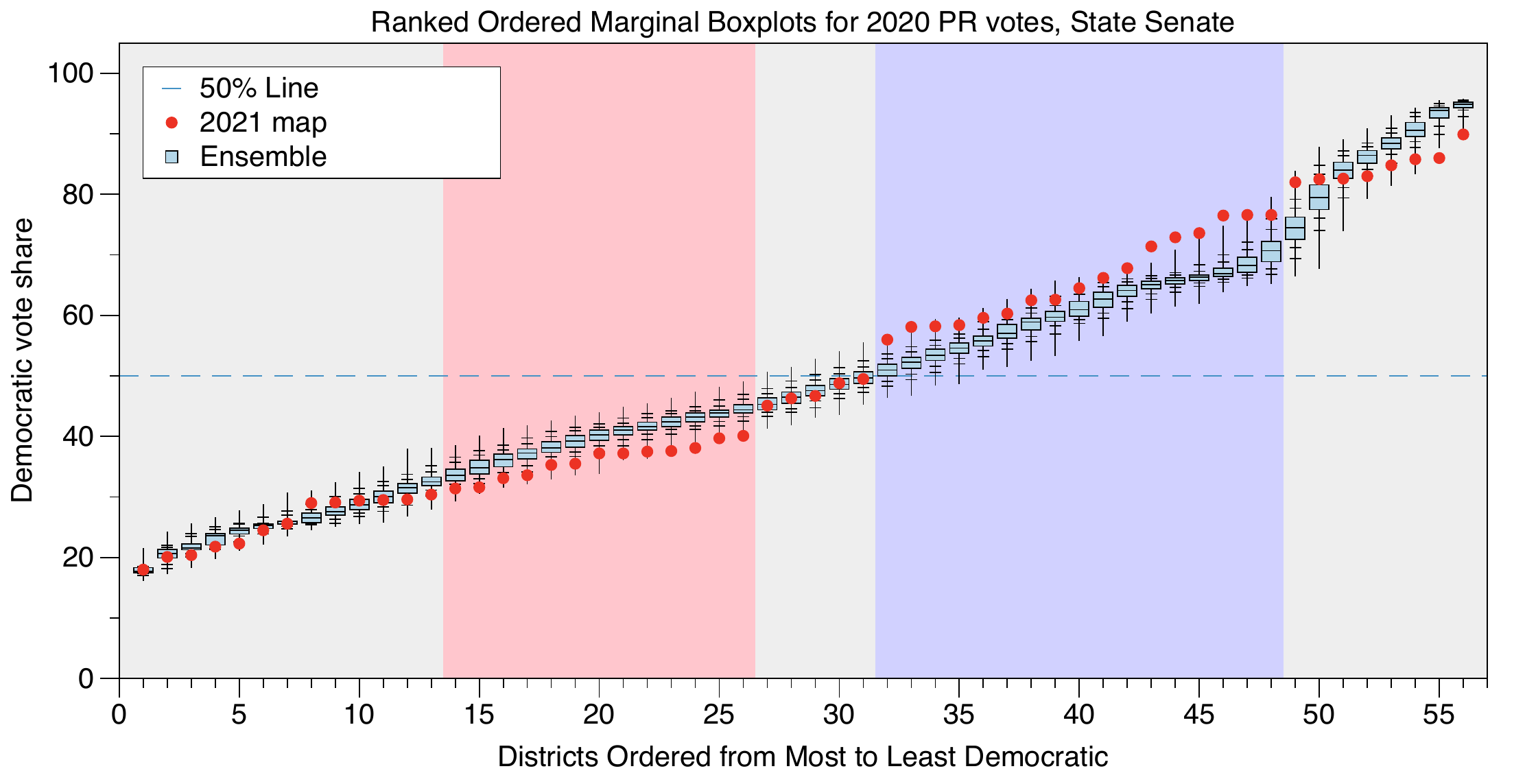}
	\end{subfigure}
	\caption{Top and middle rows: uniform swing plots for the State House (left) and Senate (right) with 2016 and 2020 presidential votes. Bottom row: rank-ordered marginal box plots for the State House (left) and Senate (right) with 2020 presidential votes.}
	\label{fig:GA_plots}%
\end{figure}

\section{The Georgia General Assembly}\label{asec:GA}
We also apply our methods to analyze the Georgia State House and Senate districting plans, which contain 180 and 56 districts respectively. As in the congressional plan, we find significant non-responsiveness caused by polarization in the 2021 enacted plans. Figure~\ref{fig:GA_plots} displays uniform swing and rank-ordered marginal box plots comparing both enacted plans to our respective ensembles.

Since these plans correspond to entire legislative bodies, rather than a single state's delegation, the critical observable is which party wins a majority of the seats (29 in the Senate and 91 in the House), rather than the raw number of seats won. The enacted plans will tend to preserve the current Republican majority in both bodies. Across all seventeen elections, Republicans win a majority of the seats even when Democrats have 54\% of the votes statewide, excepting the Senate with swings based on the 2020 presidential (29 Democratic seats) and 2020 US Senate (28 Democratic seats) elections. With Democratic vote fractions under 50\%, both enacted plans produce an unusually high number of Democrats compared to the ensemble. With Democratic vote fractions over 50\%, both enacted plans tend to elect an unusually small number of Democrats compared to the ensemble. As in the enacted congressional plan, we observe that this non-responsiveness is caused by polarization in which the more moderate rank-ordered districts are far more Republican or Democratic than is typical of plans in the ensemble.

\end{document}